\documentclass[gmd, article]{copernicus}

\usepackage{tabularx}
\usepackage{multirow}
\usepackage{amsthm}
\usepackage{amsmath}
\usepackage{float}
\usepackage{subcaption}
\usepackage{graphicx}
\usepackage[version=3]{mhchem} % Formula subscripts using \ce{}
\usepackage{hyperref}

\begin{document}

\nolinenumbers

\title{MieAI: A neural network for calculating optical properties of internally mixed aerosol in atmospheric models}

% \Author[affil]{given_name}{surname}
\Author[]{Pankaj}{Kumar}
\Author[]{Heike}{Vogel}
\Author[]{Julia}{Bruckert}
\Author[]{Lisa}{Janina Muth}
\Author[]{Gholam Ali}{Hoshyaripour}

%%\Author[]{Anika}{Rohde}

\affil[]{Institute of Meteorology and Climate Research, Karlsruhe Institute of Technology, Karlsruhe, Germany}

%% The [] brackets identify the author with the corresponding affiliation. 1, 2, 3, etc. should be inserted.

%% If an author is deceased, please mark the respective author name(s) with a dagger, e.g. "\Author[2,$\dag$]{Anton}{Smith}", and add a further "\affil[$\dag$]{deceased, 1 July 2019}".

%% If authors contributed equally, please mark the respective author names with an asterisk, e.g. "\Author[2,*]{Anton}{Smith}" and "\Author[3,*]{Bradley}{Miller}" and add a further affiliation: "\affil[*]{These authors contributed equally to this work.}".

\correspondence{Pankaj Kumar (pankaj.kumar@kit.edu)}
\runningtitle{MieAI}
\runningauthor{Kumar et al. 2023}

%\received{ 7 September 2022}
%\pubdiscuss{ 7 September 2022} %% only important for two-stage journals
%\revised{ 7 September 2022}
%\accepted{ 7 September 2022}
%\published{ 7 September 2022}

%% These dates will be inserted by Copernicus Publications during the typesetting process.
\firstpage{1}
\maketitle

\begin{abstract}
Aerosols influence weather and climate by interacting with radiation through absorption and scattering. These effects heavily rely on the optical properties of aerosols, which are mainly governed by attributes such as morphology, size distribution, and chemical composition. These attributes undergo continuous changes due to chemical reactions and aerosol micro-physics, resulting in significant spatio-temporal variations. Most atmospheric models struggle to incorporate this variability because they use pre-calculated tables to handle aerosol optics. This offline approach often leads to substantial errors in estimating the radiative impacts of aerosols along with posing significant computational burdens. To address this challenge, we introduce a novel and computationally efficient machine learning approach called MieAI. It allows for relatively inexpensive calculation of the optical properties of internally mixed aerosols with a log-normal size distribution. Importantly, MieAI fully incorporates the variability in aerosol chemistry and microphysics. Our evaluation of MieAI against traditional Mie calculations, using number concentrations from the ICOsahedral Nonhydrostatic model with Aerosol and Reactive Trace gases (ICON-ART) simulations, demonstrates that MieAI exhibits excellent predictive accuracy for aerosol optical properties. MieAI achieves this with errors well within 10\%, and it operates more than 1000 times faster than the benchmark approach of Mie calculations. Due to its generalized nature, the MieAI approach can be implemented in any chemistry transport model which represents aerosol size distribution in the form of log-normally distributed internally mixed modes. This advancement has the potential to replace frequently employed look-up tables and plays a substantial role in the ongoing attempts to reduce uncertainties in estimating aerosol radiative forcing.
\end{abstract}

%%\copyrightstatement{TEXT} %% This section is optional and can be used for copyright transfers.

\section{Introduction}  %% \introduction[modified heading if necessary]

Aerosol particles have a significant impact on Earth's radiation balance due to their interactions with solar radiation and clouds. Particles' ability to scatter and absorb radiation, known as the aerosol direct effect, is influenced by their mixing state – how different aerosol types are distributed within the population \citep{Riemer2019,Jacobson2001}. This mixing state can range from external mixing (single species) to internal mixing (mixture of species). Newly emitted aerosols usually have external mixing, while aging processes lead to internal mixing. Aerosol particles consist of diverse organic and inorganic components, showing significant variability in composition and abundance across time and space. Previous studies emphasize the importance of mixing state in understanding aerosols' optical properties (AOPs) \citep{Yao2022, Koike2014, Wang2021, Riemer2019}. For example, studies demonstrate a greater positive forcing for internally mixed black carbon aerosols under the assumption of core-shell mixing in contrast to homogeneous volume-mixing and external mixing scenarios \citep{Jacobson2001, Yu2006, Bond2006a, Zhang2008a, Moffet2009, Lack2010}.

Accurately modeling aerosol populations and predicting their impact on air quality, weather, and climate has long been a major challenge. Despite a good understanding of the underlying physics, resolving many small-scale processes, especially within atmospheric models, remains difficult. Precise quantification of AOPs, including mass extinction coefficient ($k_e$), single-scattering albedo ($\omega$), and asymmetry factor ($g$) are crucial for improving the forecasting capabilities of the atmospheric composition models (ACMs). 

Accurate representation of AOPs of internally mixed particles remains a significant challenge in ACMs \citep{Riemer2019, Brown2021}. Currently, many ACMs use large database of pre-calculated AOPs, in the form of look-up tables \citep{Dubovik2006, Ghan2006, Meng2010, gasteiger2018, Stromatas2012, Gasch2017, Ghosh2021, Geiss2022, Wang2022}. These AOPs are often archived using Mie calculations for a discrete set of chemical and micro-physical attributes (such as particle size and refractive index) \citep{Brown2021, Tuccella2020}. But aerosol size and composition have large spatial and temporal variability in model simulations. An interpolation is inevitable whenever AOPs in the model are queried for a set of input parameters different from the archived values. The interpolation may lead to non-trivial errors due to non-linearity. Such errors can be reduced by adding more parameters to the interpolation. For example, \citet{Gasch2017} implemented polynomial fits to account for the variability of median  diameter of log-normal modes during atmospheric transport (due to faster sedimentation of large particles). The database, however, grows larger as the number of parameters increases, making the offline AOPs less convenient for use with diverse applications \citep{Han2022, Geiss2022}. 

Due to the significant computational burden, online calculation of AOPs using Mie code is only feasible for specific applications and impractical for real-time use in ACMs \citep{Wang2022}. To tackle this issue, several attempts have been made for the online calculation of AOPs that often parameterize the Mie calculations for variable aerosol size and composition \citep{Ghan2001,Ghan2006,Fast2006,Klingm2014,Curci2015}. Yet, these methods are also subjected to large uncertainties and errors stemming from underlying assumptions and interpolations. This highlights the immediate demand for accurate and computationally efficient tools for online calculaton of AOPs consistent with the aerosol chemical and microphysical characteristics in ACMs \citep{Stier2007, Geiss2022}.

In recent years, the application of Machine Learning (ML) and, more specifically, Deep Learning (DL), has garnered significant prominence within the domain of weather and climate research. This prominence is reflected in its diverse applications, spanning across various aspects including weather prediction \citep{Bi2023, Zhang2023}, the refinement of numerical model outputs through post-processing \citep{Sayeed2022a}, and even the substitution of pivotal model physics \citep{Mishra2021a} and parameterizations \citep{Yuval2021,Rasp2018}. The methodologies employed encompass a variety of techniques, spanning from emulation \citep{Sharma2023} to the resolution of partial differential equations (PDEs) via widely adopted ML algorithms \citep{Huang2022,Goswami2023}. ML has undergone significant advancements in recent years, particularly after 2010, as a result of the development of effective techniques for training a neural network (NN) of considerable size. NNs excel in learning knowledge representation in very high-dimensional spaces; in forms of connecting weights in between neurons of the networks. The organisation of the networks or the network architecture is thus a mapping of the knowledge space of various domains. As demonstrated in recent studies, it is feasible to predict the optical properties of aerosol particles by means of a NN, rather than solving Maxwell's equation as in Mie calculations \citep{Lamb2023, Geiss2022, Han2022}. 

Therefore, the objective of this study is to develop a NN based emulator to replace the current aerosol optics parameterization for internally mixed aerosols used in ACMs such as ICON-ART. We present a multi-layer fully connected feed-forward NN to derive optical properties for spherical particles covering a large size range accurately and efficiently; thereby meeting the emergent requirements in both remote sensing and atmospheric modeling of aerosol particles. This study builds upon prior endeavors that employed ML techniques for emulating aerosol optics and radiative transfer modeling \citep{Belochitski2021, Mishra2021a, Ukkonen2022a, Pal2019, Lamb2023, Geiss2022, Han2022, Wang2022, Wang2023single}. The overarching objective here is to devise an approach capable of robust generalization as the existing literature lacks the discussion on the challenges of utilizing a neural network-based approach in real world applications.

\section{Methods}
\subsection{Mie Calculation of Aerosol Optical Properties}

Optical properties are a function of the particle size and the wavelength-dependent refractive indices (RIs) of the constituents of the aerosol particles \citep{Gordon2017}. Both relative RI of the particle with respect to surrounding medium and particle shape should be accounted for in radiation interaction studies \citep{Lamb2023}. If the particle shape is spherical, Mie theory can be used to calculate the optical properties. Mie theory uses Maxwell's equations to solve a 3-D electromagnetic wave equation whose solution can be written as an infinite series of products of orthogonal functions \citep{Bohren2008}. As per Mie theory, the extinction ($Q_{ext}$) / scattering ($Q_{sca}$) efficiencies and $g$ of a spherical particle can be written as:

\begin{equation}\label{eq:4}
    Q_{ext} = \frac{2}{x^2} \sum_{n=1}^{\infty}(2n+1)\mathbb{R}(a_n + b_n)
\end{equation}

\begin{equation}\label{eq:5}
    Q_{sca} = \frac{2}{x^2} \sum_{n=1}^{\infty}(2n+1)(\lvert a\rvert_n^2 + \lvert b\rvert_n^2)
\end{equation}

\begin{multline}\label{eq:6}
 g = \frac{4}{Q_{scat} x^2} \Bigg[\sum_{n=1}^{\infty} \frac{n(n+2)}{n+1} \mathbb{R}(a_n a_{n+1}^* + b_n b_{n+1}^*) \\ 
   + \sum_{n=1}^{\infty} \frac{2n+1}{n(n+1)} \mathbb{R}(a_n b_n^*)\Bigg]
\end{multline}

Here, $a_n$ and $b_n$ are the Mie scattering coefficients and $x$ is the size parameter which, in turn, is given by:

$$x = \frac{\pi d_p}{\lambda}$$ \label{eq:3}

where $d_p$ is the particle diameter. 

An approximate solution for $Q_{ext}$, $Q_{sca}$ and $g$ can be obtained by truncating the infinite series as explained by \citet{Bohren2008}. Mie codes calculate the Mie scattering coefficients ($a_n$ and $b_n$), which are solely dependent on particle diameter ($d_p$), incident wavelength ($\lambda$) and RI ($B_\lambda$), followed by determination of the number of terms required before truncation and calculation of the series. Mass extinction coefficient ($k_e$) is obtained from $Q_{ext}$ \citep{Muser2022}:

\begin{equation}\label{eq:7}
    k_e(l, \lambda, B_{\lambda}) = \frac{\int_0^\infty \frac{\pi}{4}d_p^2 Q_{ext}(d_p, \lambda, B_{\lambda}) \psi_{0,l}(d_p)dd_p}{\int_0^\infty \rho_p [\frac{\pi}{6} d_p^3] \psi_{0,l}(d_p)dd_p}
\end{equation}

where $\psi_{0,l}$ and $\psi_{3,l}$ are the parameters of the log-normal distribution for aerosol mode $l$. 

\begin{figure}[t]
    \includegraphics[width=\linewidth]{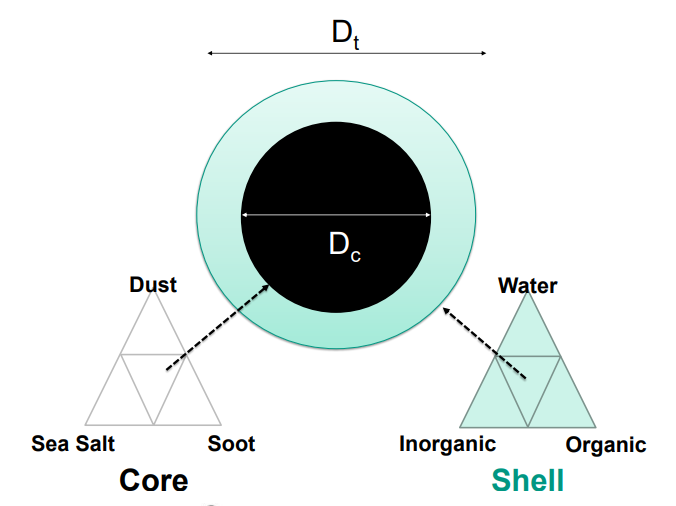}
    \caption{Coated internally mixed aerosol particle. It is assumed to be composed of a core that is insoluble and a shell that is soluble. The core consists of black carbon, volcanic ash, sea salt and dust whereas the shell consists of organic, inorganic matter (such as ammonia (\ce{NH4}), nitrate (\ce{NO3}), chlorine (\ce{Cl}), sulphate (\ce{SO4}) and sodium (\ce{Na}) and water (\ce{H2O})). Here, $D_c$ represensts the diameter of the core and $D_t$ is the total diameter of coated, mixed aerosol particle that consists of both core and shell. Refractive indices (RI) for all chemical species constituting the mixed aerosol particle except dust are obtained from \citet{Gordon2022} whereas those for dust are obtained from \citet{DiBiagio2019}.}
    \label{fgr:example2}
\end{figure}

To calculate the optical properties of the internally mixed aerosol particle using Mie calculations, some assumption are required. Mie theory assumes that the particles have spherical shapes.  In reality, the majority of aerosol particles are non-spherical. However, the process of liquid coating frequently leads to the formation of spherical coating surfaces, thus justifying the assumption of particle sphericity in mixed mode models. Recent studies suggest that coated particles can also exhibit non-spherical shapes, which complicates this assumption \citep{li2009observation, chakrabarty2018, fierce2020, Wang2022, kelesidis2022} . Nevertheless, the use of coated spheres remains a practical approximation in many cases and is widely used configuration in ACMs \citep{ Ma2012, Muser2020, Geiss2022, Wang2022}. In this study, aerosol particles were assumed to be spherical in a core-shell configuration, with solid phase as the core and liquid species as the shell. Both core and shell are considered as ternary systems of different chemical species. For example, core is the ternary system consisting of dust, sea salt and soot whereas the shell is constituted by water, inorganic and organic species as shown in Fig.~\ref{fgr:example2}. This assumption does not imply the one-to-one existence of such mixtures in nature. Rather, it covers a wide range of the possible RIs for core and shell accruing in the atmosphere which is shown in Fig.~\ref{fig:figures1}. We employ the PyMieScatt Python library for computing the optical characteristics of a coated sphere using Mie theory \citep{Sumlin2018}. This library is built on Mie codes originally written by \citet{Matzler2002} and \citet{Bond2006}, rooted in the concepts presented by \citet{Bohren2008}. 

\begin{figure}[t]
%\begin{subfigure}[t]{\linewidth}
%    \centering\includegraphics[width=0.95\linewidth]{./figures/MieAI.PNG}
%    \caption{A schema of the proposed ML based approach for accelerating the estimation of aerosol optical properties in CTMs}
%    \label{fgr:example1}
%    \vspace*{6mm}
%\end{subfigure}

%\begin{subfigure}[b]{\linewidth}
    \centering\includegraphics[width=0.95\linewidth]{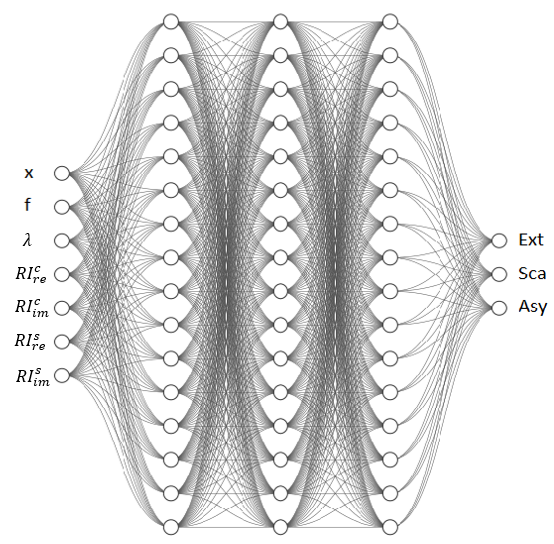}
%    \caption{MieAI Architecture}
%    \label{fgr:example31}
%\end{subfigure}

\caption{MieAI Architecture. MieAI is a NN based model with multiple hidden layers. The first and last layers represent input and output of MieAI respectively. Here, size parameter (x), wavelength ($\lambda$), coating fraction (f), real and imaginary parts of refractive indices for both core ($RI^c_{re}$ and $RI^c_{im}$) and shell ($RI^s_{re}$ and $RI^s_{im}$) constitute the input of MieAI whereas the extinction (Ext), scattering (Sca) Efficiency and asymmetry parameter (Asy) are the output.}
\label{fig:figures0}

\end{figure}

\subsection{Emulation of the Mie Calculation: MieAI}

In this study, we propose a multi-layer fully connected NN popularly known as multi-layer perceptron (MLP) to emulate the calculation of AOPs using Mie calculation i.e. MieAI. As a universal function approximator, the feed-forward NN is ideally suited for modeling nonlinear processes. The schematic diagram of the MLP is shown in Fig.~\ref{fig:figures0}. Specifically, it is used here to establish the relationships between the micro-physical parameters of aerosol particles and corresponding single-scattering properties \citep{Chen2022, Wang2022, Han2022, Geiss2022}. Its feature is the interconnection of neurons with all nodes in the front and rear hidden layers. The output $O^{(l)}_i$ of the $i$-th node in the fully connected layer $l$ can be calculated from the output of the previous layer $l - 1$ with a non-linear activation function ($\phi$).

\begin{equation} \label{eq:8}
    O^{(l)}_i =  \phi \Bigg(\sum^{N^{(l-1)}}_{j=1} w^{(l)}_{i,j}O_{j}^{(l-1)} + b^{(l)}_{i}\Bigg)
\end{equation}

Here, $w^{(l)}_{i,j}$ represents the weight of the $j$-th neuron in the layer $l-1$ to the $i$-th neuron in the layer $l$ and $b^{(l)}_{i}$ represents the bias term of the $i$-th neuron in the layer $l$. $N^{(l-1)}$ is the number of neurons in layer $l-1$. 

For estimating AOPs using MieAI, 7 aerosol micro-physical parameters are regarded as input features (X = [ $x_1$, $x_2$, …, $x_{7}$ ]) and 3 single-scattering properties (Y = [ $Q_{ext}$, $Q_{sca}$, g ]) are output targets as shown in Fig.~\ref{fig:figures0}. Here the input features are the size parameter ($x$), wavelength ($\lambda$), coating fraction ($f$) for coated, internally mixed aerosol and RIs for both core ($RI^c$) and shell ($RI^s$). Using a dataset comprising known input and output matrices, denoted as X and Y respectively, the model undergoes training to optimize its parameters – weights ($w$) and biases ($b$). This optimization is achieved via back-propagation, which minimizes the cost function $C_y$:

\begin{equation} \label{eq:9}
    C_y = \sum_{i=1}^{N} (y_{true} - y_{pred})^2
\end{equation}

This function quantifies the error between the predicted values ($y_{pred}$) generated through forward propagation in the NN and the actual values ($y_{true}$). The cost function $C_y$ is differentiable with respect to the model parameters ($w$ and $b$), enabling the application of various gradient descent techniques for efficient optimization.

\subsection{Training Data and its preprocessing}

To facilitate MieAI training, a total of 30 distinct combinations of core and shell chemical compositions are considered, as outlined comprehensively in Table~\ref{tbl:s1}. The computation of optical characteristics relies on wavelength-dependent RI. As emphasized by \citet{Rieger2017}, distinct peaks in the real component of the RI manifest as prominent maxima in $Q_{ext}$. Simultaneously, the $\omega$ and, consequently, the absorption efficiency ($Q_{abs}$) is governed by the imaginary component of the RI. In Fig.~\ref{fgr:ri}, we present the real and imaginary components of RI for the chemical species composing both the core and shell of aerosol particles \citep{Rieger2017, Muser2020, Hoshyaripour2019}. Fig.~\ref{fgr:example41} illustrates the variations in the real and imaginary components of the RI for internally mixed and coated particles as a function of changes in the chemical composition of the core and shell across various wavelengths of solar radiation. The real part of the RI exhibits a range from 1.1 to 2.75 for the core and 1.2 to 2 for the shell, contingent upon the specific chemical compositions of the core and shell. Meanwhile, the imaginary component varies from values as low as $10^{-8}$ to 0.5 for the core and from $10^{-9}$ to 1 for the shell. It's noteworthy that the core is characterized as a volume-averaged ternary system involving mineral dust, sea salt, and soot, while the shell is likewise modeled as a ternary system, featuring water, inorganic, and organic constituents. 

\begin{figure}[t]
    \includegraphics[width=\linewidth]{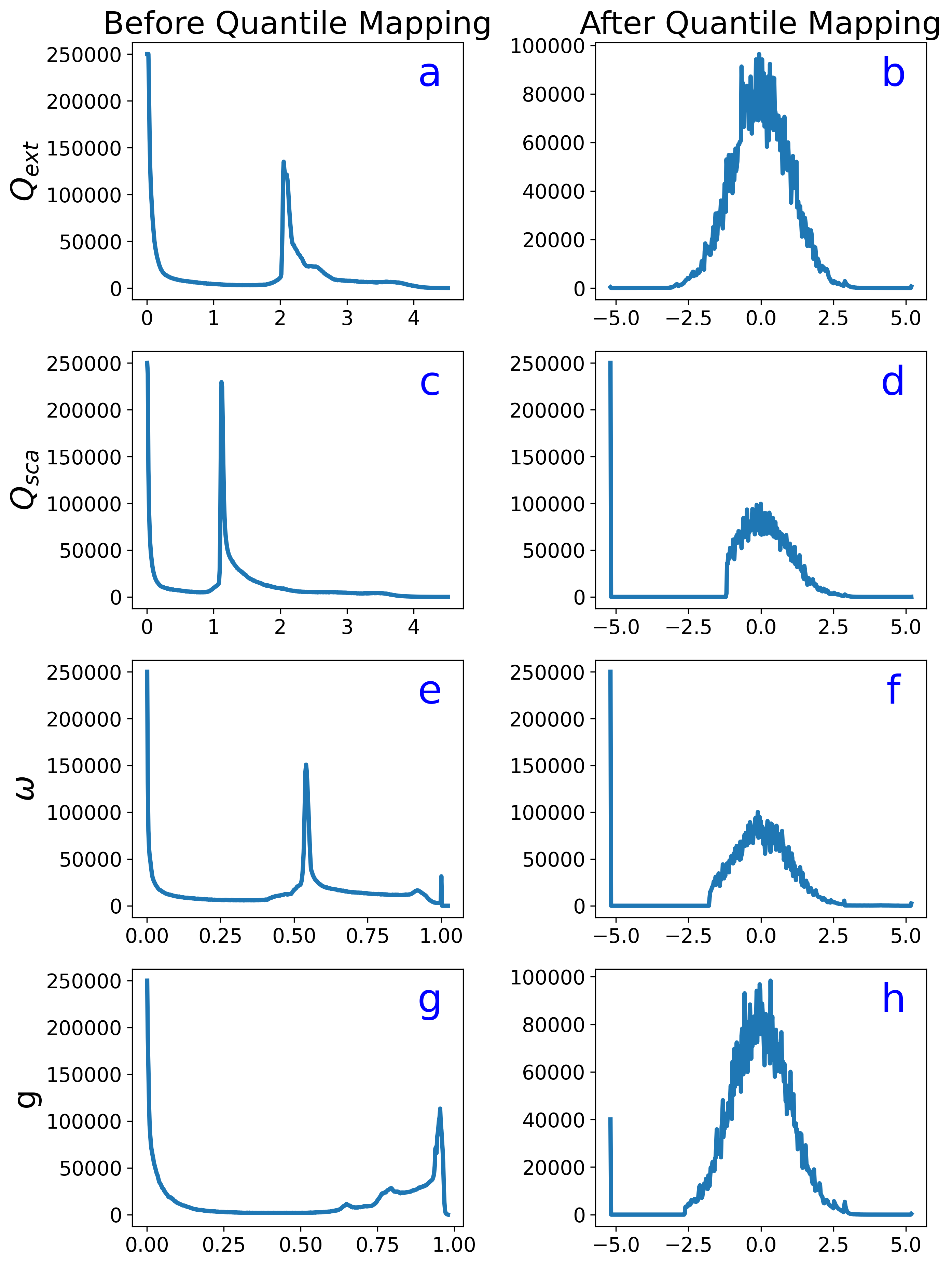}
    \caption{Variation of target AOPs before and after quantile mapping. Quantile mapping transforms input features to a Gaussian distribution with mean 0 and standard deviation 1.}
    \label{fgr:mapping}
\end{figure}

The training, test and validation datasets for MieAI are generated by randomly selecting 600,000 samples (about 2 \%) from more than 45 million possible combinations of input features arising from varying wavelength (0.2 to 100 $\mu$m), shell thickness (from 0 to 40\% of total diameter with a step wise of 0.1\%), core diameter (from 10 nm to 20 $\mu$m) and RI by considering 30 different combinations for core and shell as discussed before. Randomly selected samples were divided into training (70\%), validation (15\%) and test datasets (15\%) while optimising the NN architecture and parameters.

Both input and target datasets have a large variability; hence it is important to normalize them before feeding to NN for training in order to improve the model learning ability. Hence, input and target data to NN model is transformed using Min-Max normalization before being fed to NN model. Afterwards, the output from the NN model is denormalized to its original optical properties space. We first normalized the training dataset and used the same normalization scale to transform validation and test datasets to avoid data leakage during model training.

Due to the non-normal distribution of the target AOPs in training dataset, we perform a quantile distribution mapping over the raw target AOPs to a normal distribution. Quantile mapping transforms all input features to the same target distribution (Gaussian distribution in this case) based on the formula $G^{-1}(F(X))$ where $F$ is the cumulative distribution function (CDF) of the input feature and $G^{-1}$ is the quantile function of the target distribution G \citep{Jakob2011}. Quantile mapping smooths out uneven distribution and is influenced less by outliers unlike scaling methods like min-max transformation. Quantile mapping has been used extensively in meteorology for bias correction \citep{Hertig2019} and statistical downscaling \citep{Abatzoglou2012}. We use the python library scikit-learn for performing quantile mapping in this study. As shown in Fig.~\ref{fgr:mapping}, the raw training dataset for $g$ is bi-modal with one peak over 0 and another over 1. While non-linear algorithms like MieAI may not have a Gaussian distribution assumption, however they perform better if variables have a Gaussian distribution. Thus, mapping to the normal distribution improves the generalization of the trained network. During inference, the predicted AOPs are transformed back to the original distribution using inverse quantile transform with the same parameters used during the training.

\subsection{Optimisation and assessment of MieAI}

In addition to the model parameters optimized by the NN training procedure, there are hyper-parameters that define the model architecture and control the learning process, such as the number of hidden layers, the number of neurons in each layer, the activation function, batch size and the learning rate of the optimizer. The mean squared error (eq. \ref{eq:9}) is employed as the loss function for the optimizer to minimize.  We apply a non-linear activation function to all of the layers except the output where we apply linear activation to restrict the NN output between 0 and 1 \citep{Cachay2021}. After the training, the weight matrices in the NN are saved and used afterwards for evaluation using ICON-ART simulations. 

To assess the performance of the network, we used the coefficient of determination ($R^2$) and Mean Absolute Percentage Error (MAPE) as metrics to evaluate the fitness of the predictions with the true values. $R^2$ is defined as:

\begin{equation}\label{eq:10}
    R^2 = 1 - \frac{\Sigma^M_{i=1} (y_i - f_i)^2}{\Sigma^M_{i=1} (y_i - \Bar{y})^2}
\end{equation}

Here, $f_i$ is the value predicted by MieAI and $y_i$ is the true value. $\Bar{y}$ is the average of all true values. The closer $R^2$ is to 1, the higher the performance of MieAI. The MAPE metrics is defined as:

\begin{equation}\label{eq:11}
MAPE = \frac{100}{N} \sum^N_{i=1} \bigg| \frac{Y_{mie} - Y_{MieAI}}{Y_{mie}} \bigg|
\end{equation}

Here, $Y_{MieAI}$ is AOP prediction from MieAI, $Y_{mie}$ is the AOP estimated using Mie theory and N is the number of times AOPs are predicted using MieAI.

\begin{table}[t]
\caption{Hyper-parameter tuning of MieAI model. We use the \textit{keras-tuner} hyper-parameter optimization library to tune the hyper-parameters of MieAI.}
\begin{tabular}{lccc}
\tophline
& Tried value ranges & Best value \\
\middlehline
Number of Hidden Layers & 1--5 & 4 \\
Neurons per Hidden layer & 8, 16, 32,64 & 64 \\
Minibatch Size & 128, 256, 512, 1024 & 128 \\
Activation Function & relu, gelu, sigmoid, tanh & gelu \\
Optimizer & adam, sgd, rmsprop & adam \\
Initial Learning rate & $10^{-1}$, $10^{-2}$, $10^{-3}$, $10^{-4}$ & $10^{-2}$ \\

\bottomhline
\end{tabular}
\label{tbl:1}
%\belowtable{} % Table Footnotes
\end{table}

 To avoid over-fitting and other training related issues, we chose our NN hyper-parameters using \textit{keras-tuner} hyper-parameter optimization library and apply early stopping with patience parameter set as 50 \citep{Shawki2021}. The hyper-parameters of the model have been meticulously optimized through the application of Bayesian optimization. The entire hyper-parameter tuning procedure is executed in a two-stage approach, wherein each stage serves to fine-tune distinct aspects of the model. In the first stage, we focus on optimizing critical architectural components, including the number of hidden layers, the neuron count in each hidden layer, activation functions, and the choice of optimizer. Subsequently, the second stage hones in on further enhancements by fine-tuning the learning rate and the batch size of the training data, for the NN selected in the first stage. During hyper-parameter optimization, we trained various NN architectures for 200 epochs. The corresponding MSE values for these diverse NN architectures are presented in Table~\ref{tbl:s2} (first stage) and Table~\ref{tbl:s3} (second stage). The resultant optimal values for all hyper-parameters are shown in Table~\ref{tbl:1}. 
 
 As depicted in Table~\ref{tbl:s2}, the MieAI model with Adam optimizer having 5 hidden layers with 64 neurons in each layers and GELU activation function performed the best with the lowest MSE. With the aim to select the most accurate NN with smallest possible number of trainable parameters, we performed the second stage of tuning wherein we varied the number of hidden layers and the number of neurons in each layer along with the batch size and learning rate of Adam optimiser selected after the first stage tuning. As shown in Table~\ref{tbl:s3}, the MieAI model with 4 hidden layers outperformed the 5 layer NN as selected in first stage when batch size and learning rate were also optimised. We apply early stopping with patience set as 50 and reduce the learning rate of the optimizer by one-fifth if the validation loss plateaus during both hyper-parameter tuning and training of the network. Therefore, the best NN after hyper-parameter optimisation is a MLP with 4 hidden layers each having 64 neurons trained using Adam optimizer with learning rate of 0.01 and training batch size of 128. 

\subsection{ICON-ART model system}

In addition to evaluating the trained MieAI using test datasets, we conducted three reference ICON-ART simulations for real-world events to validate the MieAI prediction of AOPs against Mie calculations. The ICON modelling framework excels in solving the nonhydrostatic and compressible Navier-Stokes equations on an icosahedral-triangular grid \citep{Zangl2015}. This model exhibits versatility in predicting various processes across scales, from global to local, as highlighted by \citet{Giorgetta2018} and \citet{Heinze2017}. Complementing the ICON model, the ART module forms an integral part responsible for simulating trace gases and aerosols in both the troposphere and stratosphere. This module encompasses processes spanning emission, transport, physicochemical transformation, removal of gases and aerosols as well as their interactions with clouds and radiation \citep{Rieger2015, Gasch2017, Schroter2018}. Deutscher Wetterdienst (DWD) uses ICON and ICON-ART for operational weather and mineral dust forecasting and pollen, respectively.  

ICON-ART uses the European Centre for Medium-Range Weather Forecasts (ECMWF) radiation scheme \textit{ecRad} \citep{Hogan2018} as the standard radiation scheme for numerical weather prediction \citep{Rieger2019, Seifert2023}. To calculate the local radiative transfer parameters, \textit{ecRad} needs the $k_{e l,j}$, $\omega_{l,j}$ and $g_{l,j}$ for every mode $l$ and every waveband $j$ for 30 wavelength bands between 0.2 and 100 µm. These are often obtained using Mie calculations. Together with the local aerosol mass mixing ratios ($\psi_{3,l}$) from ART and air density ($\rho_{a}$), they allow for calculation of the volume specific extinction coefficient \citep{Muser2022}:

\begin{equation}\label{eq:1}
    \beta_{ext, l,j} = k_{e l,j} \cdot \rho_{a} \cdot \psi_{3,l} \cdot 10^{-6}
\end{equation}

$\omega$ gives the scattering coefficient:

\begin{equation}\label{eq:2}
    \beta_{scat, l,j} = \omega_{l,j} \cdot \beta_{ext, l,j}
\end{equation}

These volume specific properties are then converted to values per model layer by multiplying with the respective layer height ($\Delta z$), followed by summation across all model layers to calculate total aerosol optical depth (AOD) for the ART aerosol within a specific waveband. These computed values then serve as input parameters for the radiation scheme \citep{Gasch2017}. This approach ensures full coupling and feedback between aerosol processes, radiation, and the atmospheric state \citep{Hoshyaripour2019, Shao2011}.

The present study focuses on the interaction of internally mixed aerosols with radiation, which is comprehensively addressed through the use of the AEROsol DYNamic module (AERODYN) in ICON-ART. This module enables examination of aerosol dynamics processes, including nucleation, condensation and coagulation that generate internally mixed aerosols. AERODYN comprises flexible number of log-normal modes (up to 10) that accounts for Aitken, accumulation, and coarse particles in soluble, insoluble, and mixed states, alongside a giant insoluble mode \citep{Muser2020}. The term "mixed state" here pertains to an aerosol that is composed of an insoluble core and a soluble shell, and the latter constitutes no less than 5\% of the overall mass of the aerosol. The prognostic equations for number density and mass concentration are solved for each species and each mode while maintaining constant standard deviations. There exist two distinct circumstances that result in the alteration of particle modes. The first circumstance is when the mass threshold of soluble coating on insoluble particles surpasses 5\%, leading to a transition from insoluble to mixed mode. The second circumstance is when the diameter threshold of the soluble and mixed mode is exceeded, resulting in a shift to a larger mode. Alterations in the particle modes can modify the optical properties of particles with consequential impacts on both the atmospheric state and radiation \citep{Muser2020, Bruckert2023}. 

In ICON-ART, each aerosol component was assigned a RI, and the RI values were obtained from \citet{DiBiagio2019} for dust and \citet{Gordon2022} for other species. The volume-average mixing rule is used to compute the complex RI of both core and shell, which then serves as input for the core–shell calculation. To facilitate a comparison between Mie calculations and MieAI predictions, we initially derived bulk AOPs for each aerosol mode by aggregating optical properties across individual aerosol population bins. To achieve this, we initially mapped each aerosol mode, based on its median diameter, to 15 log-normal bins, as illustrated in Fig.~\ref{fgr:example3}. Both Mie calculations and MieAI emulation were then applied to these bins, and the results were subsequently integrated to obtain bulk optical properties for each mode. For our validation, we employed RI values at a wavelength ($\lambda$) of 550 nm.

\subsection{Case Studies}
In order to validate accuracy and computational efficiency of MieAI, we apply both MieAI and Mie code to the outputs of three different case studies with different aerosol species. In the following, we briefly explain the experiments. Table ~\ref{tbl:3} summarizes the relevant aerosol characteristics in each experiment. It is noteworthy that MieAI was exclusively trained on a dataset featuring shell thicknesses up to 40\%, while the comparisons in all three cases encompass shell thicknesses beyond 40\%, reaching up to 50\%. This extension aims to demonstrate the generalization capability of MieAI. Additionally, it is imperative to recognize that the stability of the Mie code output diminishes as the coating exceeds 50\%. Furthermore, we hypothesize that particles undergo a transition into an optically soluble mode beyond a coating threshold of 0.5 i.e. they are treated as particles in soluble mode instead of the mixed mode. Importantly, our focus is not to validate the model simulations in these events. Rather, we aim at evaluating the MieAI performance with real model data. 

\begin{table*}[thbp]
\caption{Summary of the mixed mode properties in case studies.}
\begin{tabular}{lrccc}
\tophline
Case & Bin diameter range (nm) & Core components & Shell components \\
\middlehline

2021 La Soufrière eruption & 10 -- 1200  & Volcanic ash & $H_2O/SO_4^{2-}$ \\
2019--20 Australian Wildfire & 3 -- 1000  & Soot & $H_2O/SO_4^{2-}/NO_3^{-}$ \\
Summer 2019 dust event & 100 -- 5000 & Dust, sea salt and soot & $H_2O/SO_4^{2-}/NO_3^{-}/NH_4^{+}$\\

\bottomhline
\end{tabular}
\label{tbl:3}
%\belowtable{} % Table Footnotes
\end{table*}

\subsubsection{2021 volcanic eruption of La Soufrière}

The first numerical experiment (case volcano) is a simulation of the last La Soufrière eruption in April 2021 and was performed by \citet{Bruckert2023}. Located on the St Vincent island in the Caribbean, the La Soufrière volcanic eruption occurred during 09--21 April 2021 and emitted volcanic material such as ash and SO$_2$ in 49 eruption phases. The simulation covered the initial four days of the eruption, encompassing 43 of the 49 eruption phases starting from 09 April at 12 UTC. The simulation had a grid spacing of 13 km with 2 nested grids around the volcano with 6.6 and 3.3 km grid spacing, respectively. The model employed 90 vertical levels to resolve the atmosphere up to 75 km. The experiment accounts for aging of volcanic ash (ash coated by sulfate-water mixture) due to aerosol dynamics. More details on this experiment is provided by \citet{Bruckert2023}. 
%In the global domain and first nest, convection was parameterized with the Tiedke-Bechtold scheme \citep{Tiedtke1989,Bechtold2008} whereas the innermost nest resolves convection. For the volcanic emission of SO$_2$ and ash, the eruption phases were resolved using the approach adopted by \citet{Bruckert2022}. This approach calculates the mass eruption rate with the 1D volcanic plume rise model FPlume based on the plume heights from \citet{Horvath2022} and distributes the mass of observed SO$_2$ over the phases. The experiment accounted for the interaction of aerosols and radiation, but not aerosol-cloud interaction. Additionally, it considered a simplified OH-chemistry \citep{weinmer2017} and included aerosol dynamics using the AERODYN module.

\subsubsection{2019--20 Australian wildfire}

The second case (case wildfire) investigated the catastrophic 2019--20 Australian wildfires in Queensland, which severely affected over 7.5 million hectares and caused a decline in air quality. A 1-day simulation was performed on November 23rd in an area on the eastern coast of Australia (150°E-160°E, 23°S-33°S). The model featured a grid spacing of 6.6 km, extending vertically to 20 km with 125 levels in a limited area setting. The emission fluxes are taken from the Global Fire Assimilation System (GFAS). 25\% of the particle mass is emitted in the Aitken mode and 75\% in the accumulation mode. The emission height is parameterized with the plume rise model according to \citet{Freitas2005,Freitas2007, Freitas2010,Walter2016}.

%and employed the same convection scheme as in case volcano i.e.  the Tiedtke-Bechthold scheme. The emission flux for biomass burning aerosols is taken from the Global Fire Assimilation System (GFAS), 25\% of the particle mass is emitted in the Aitken mode and 75\% in the accumulation mode. The emission height is parameterized with the plume rise model according to \citet{Freitas2005,Freitas2007, Freitas2010,Walter2016}. The chemical tracers (such as CH$_4$, C$_2$H$_6$, C$_3$H$_8$, CH$_3$COCH$_3$, CO, NH$_3$, NO$_2$, SO$_2$, DMS, HNO$_3$) are initialized with CAM-Chem data \citep{Emmons2020}. Akin to case volcano, we employed a simplified OH-chemistry mechanism and enabled new particle formation via nucleation along with particle coagulation and gaseous species condensation using the AERODYN module.

\subsubsection{Summer 2019 dust event over central Europe}

The third case study (case dust) centered around a dust event over central Europe during 22--27 June 2019, involving global-scale simulations with a 40 km grid. The simulation considers comprehensive aerosol emissions (including sea salt, dust, and soot) and their dynamic processes (such as nucleation, condensation and coagulation), with simplifications made in gas-phase chemistry for operational forecasting. Similar to the wildfire case study, chemical species were reinitialized daily using CAM-Chem data.

\section{Results}
\subsection{MieAI training and testing}
\begin{figure}[t]

 \begin{subfigure}[c]{\linewidth}
    \centering\includegraphics[width=\linewidth]{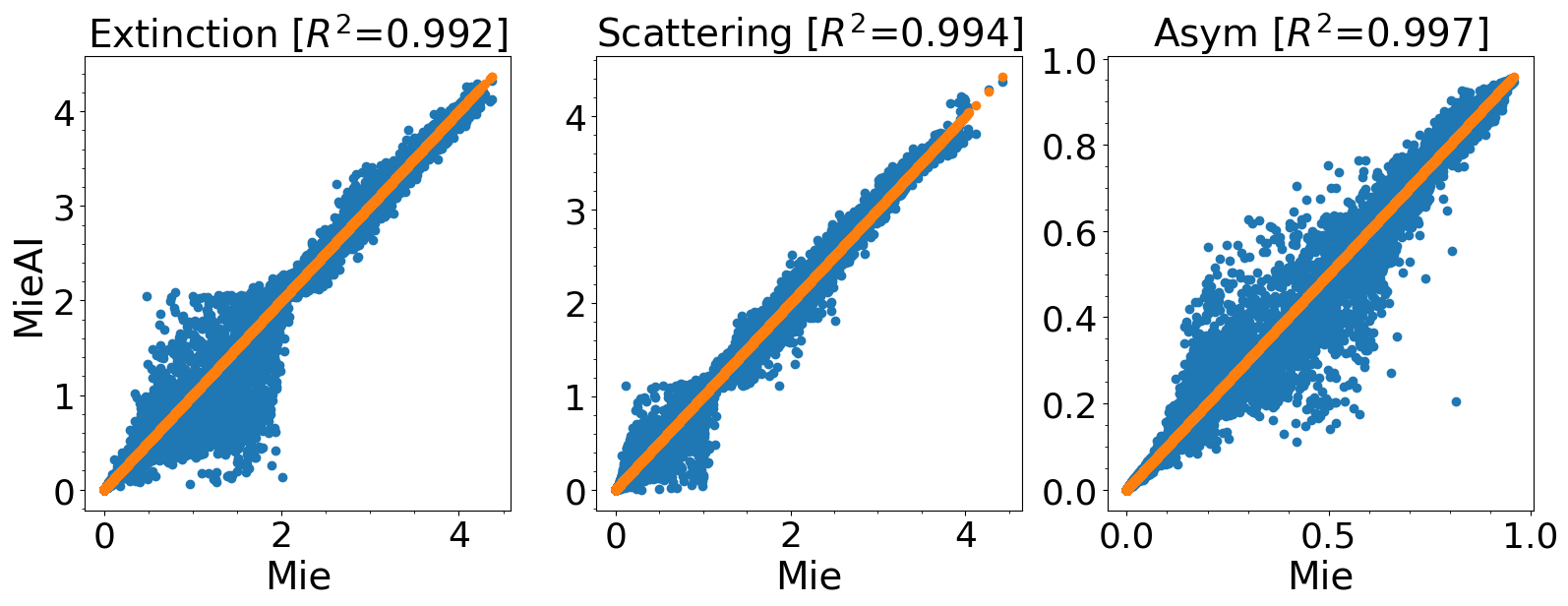}
    \caption{MieAI predictions against true AOPs estimated using Mie calculations for the test dataset.}
    \label{fgr:example5}
    \vspace*{5mm}
\end{subfigure}

\begin{subfigure}[b]{\linewidth}
    \centering\includegraphics[width=\linewidth]{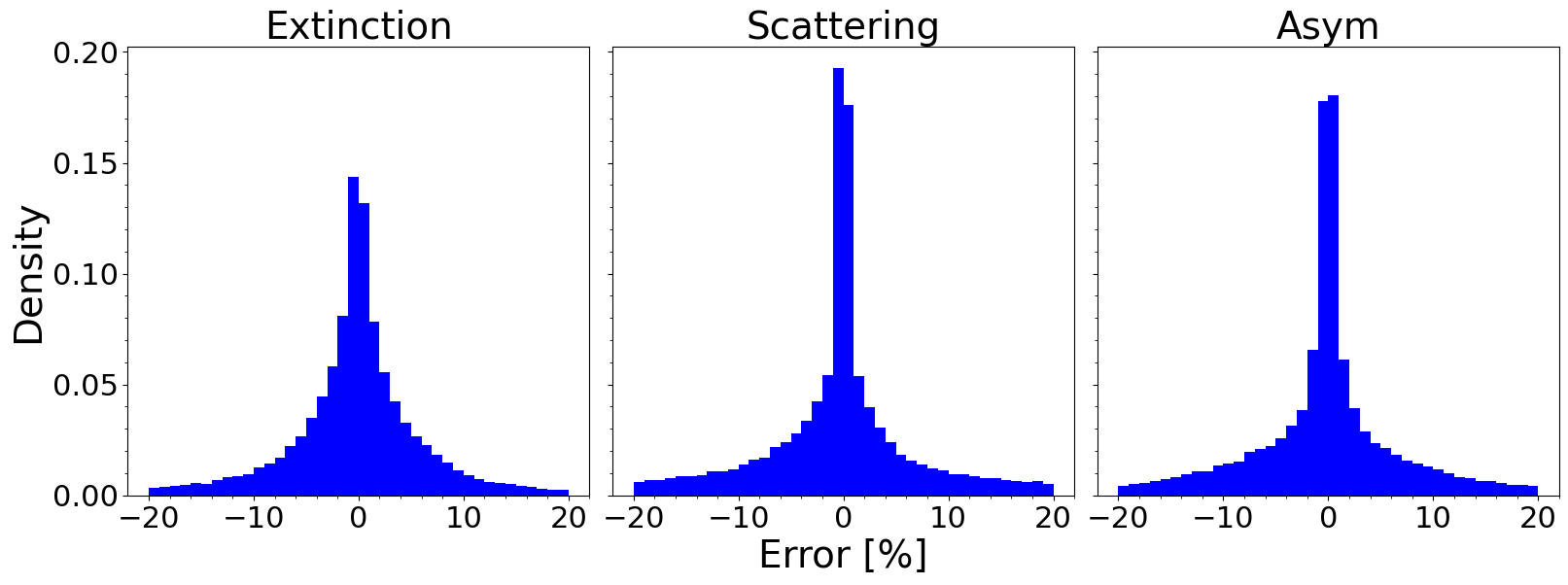}
    \caption{Distribution of NN errors for the test dataset. Here, error reported is the percentage error of MieAI with respect to true AOPs estimated using Mie calculations.}
    \label{fgr:example6}
    \vspace*{2mm}
\end{subfigure}

\caption{MieAI training and evaluation.}
\label{fig:figures2}
\end{figure}

In this study, we use a MLP with multiple hidden layers (named MieAI) to emulate the calculation of AOPs using Mie theory. MieAI considers the mixing state of particles by incorporating inputs such as size parameter, shell thickness, and RI of both core and shell. It then outputs three AOPs, including $Q_{ext}$, $Q_{sca}$, and $g$. $\omega$ is calculated from $Q_{ext}$ and $Q_{sca}$ using eq.~\ref{eq:2}. 

MieAI model, selected after hyper-parameter tuning, is trained for 5000 epochs until the loss function is minimized, resulting in an optimized network. The NN was trained on 500,000 training data samples whereas verification was done on 100,000 test samples randomly chosen from 600,000 Mie samples. Fig.~\ref{fgr:example4} provides a comprehensive visualization of the dynamic evolution of two crucial loss metrics, the MSE and the MAPE, throughout the training process of our NN model. These loss metrics are pivotal for assessing the model's performance, particularly in its capacity to accurately approximate AOPs. The observed trends in this figure offer profound insights into the model's convergence and its effectiveness in learning from the training data. Notably, the continuous and monotonic decrease in validation losses, both in terms of MSE and MAPE, serves as a strong indicator of the model's robust fitting to the data. This persistent reduction in validation losses underscores the model's consistent improvement in its ability to predict AOPs accurately as the training progresses. Such a trend is highly promising, as it demonstrates the model's capacity to continually refine its representations and effectively grasp the intricate relationships inherent in the AOP data. 

It's important to highlight that our model's training incorporates an early stopping mechanism, with a patience parameter set at 50. This strategy ensures that the model training halts at the 2548th epoch, optimizing the MieAI model with a validation MSE of 0.01187. This early stopping mechanism is a prudent approach to prevent overfitting and ensure that the model generalizes well to unseen data.

To verify the optimized network, we evaluated its performance by comparing its AOP predictions against the true AOP values estimated using Mie calculations and the $R^2$ values for different AOPs modelled in this study are shown in Fig.~\ref{fgr:example5}. As depicted in the figure, our trained NN model demonstrates a commendable ability to model AOPs effectively, as evidenced by high $R^2$ values of 0.994, 0.994, and 0.997 for $Q_{ext}$, $Q_{sca}$, and $g$, respectively. These results underscore the robust learning capability of the selected NN model, affirming its aptitude for capturing the intricate relationships within the data.

However, it is worth noting that while MieAI excels in predicting these three AOPs overall, there are specific regions where it encounters challenges. In particular, these challenges become apparent in the case of $Q_{ext}$ and $Q_{sca}$, especially when these values fall below 2 and 1, respectively. In these regions, MieAI appears to struggle, leading to more substantial discrepancies between its predictions and the actual values.

To gain deeper insights into these discrepancies, we examine the distribution of relative errors in MieAI predictions, as illustrated in Fig.~\ref{fgr:example6}. This analysis reveals that MieAI tends to underestimate $Q_{ext}$ and $Q_{sca}$ slightly when compared to those calculated using the Mie theory. However, notably, there is no such bias observed in the case of $g$. These findings provide valuable insights into the performance characteristics of MieAI and highlight specific areas where further model refinement may be warranted.

\subsection{MieAI validation using ICON-ART simulations}

\begin{figure*}[t]
    \includegraphics[width=0.88\linewidth]{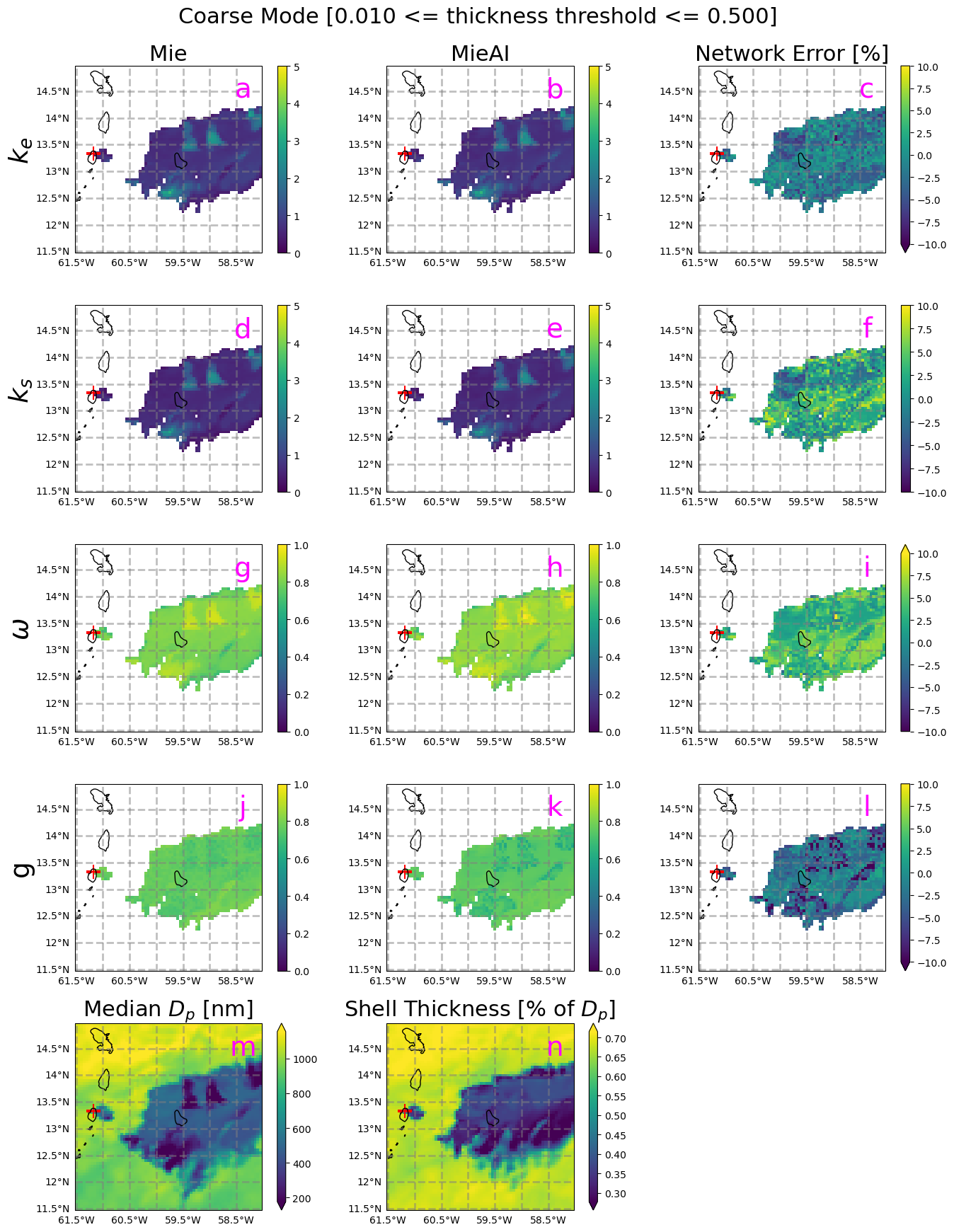}
    \caption{Comparison of AOPs predicted by MieAI against those estimated using Mie theory for coarse mode internally mixed aerosol particles at an altitude of 15 km above sea level for the La Soufrière volcanic eruption (denoted by the plus symbol) event simulated using ICON-ART. Here, left column (a, d, g, j) shows the AOPs estimated using Mie theory, middle column (b, e, h, k) shows the same predicted from MieAI and right column (c, f, i, l) shows the relative error of MieAI AOPs prediction against Mie calculations. Panel (m) shows the geographical distribution of the aerosol median diameter simulated using ICON-ART whereas Panel (n) shows the geographical variation of shell thickness, as a fraction of the total particle diameter (in percentage), of the coated aerosol.}
    \label{fgr:example7}
\end{figure*}

\begin{figure*}[t]
    \includegraphics[width=\linewidth]{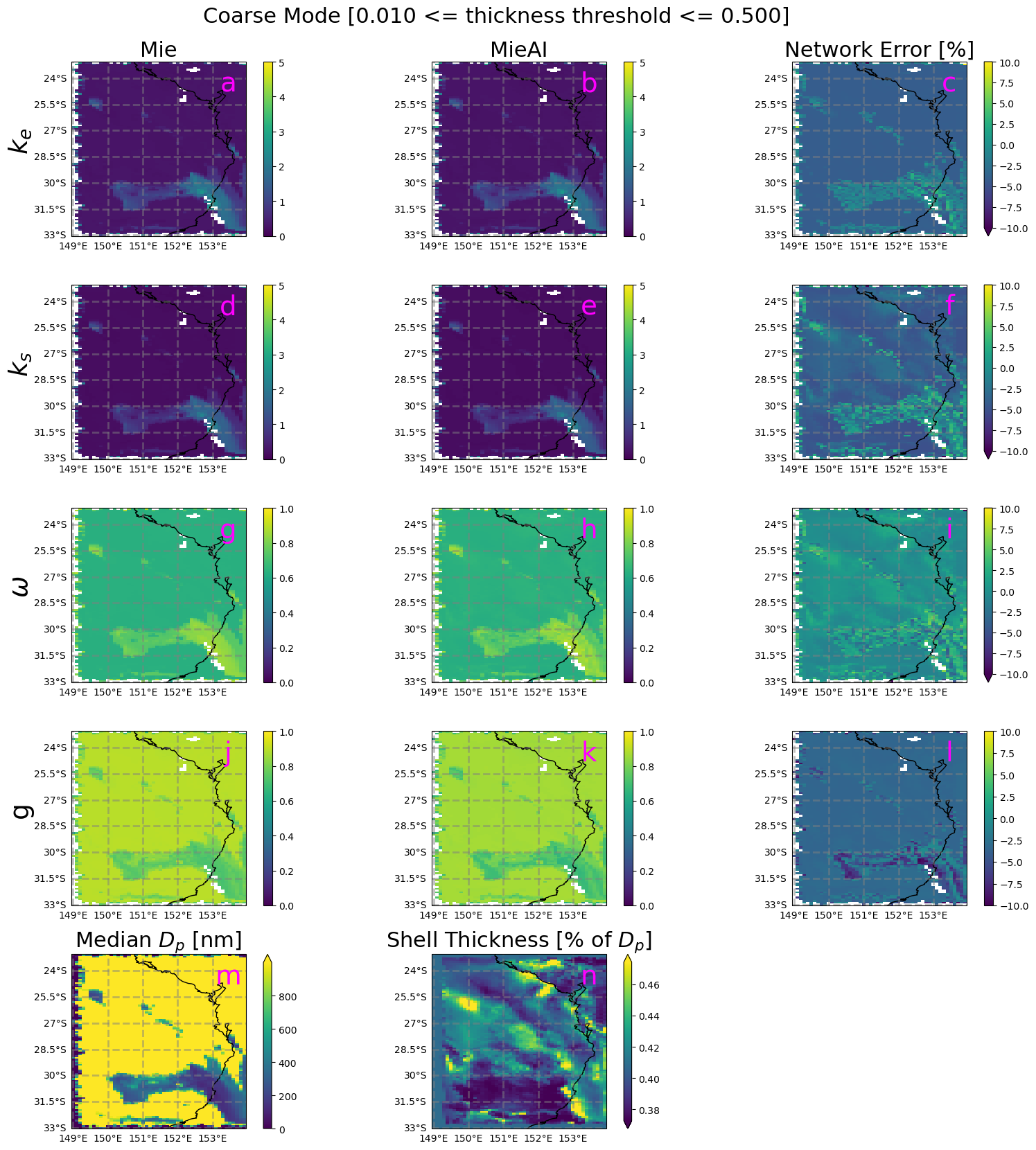}
    \caption{Same as Fig.~\ref{fgr:example7} but for coarse mode internally mixed aerosols at an altitude of 850 m above sea level in ICON-ART simulation of Australian biomass burning event.}
    \label{fgr:example8}
\end{figure*}

\begin{figure*}[t]
    \includegraphics[width=\linewidth]{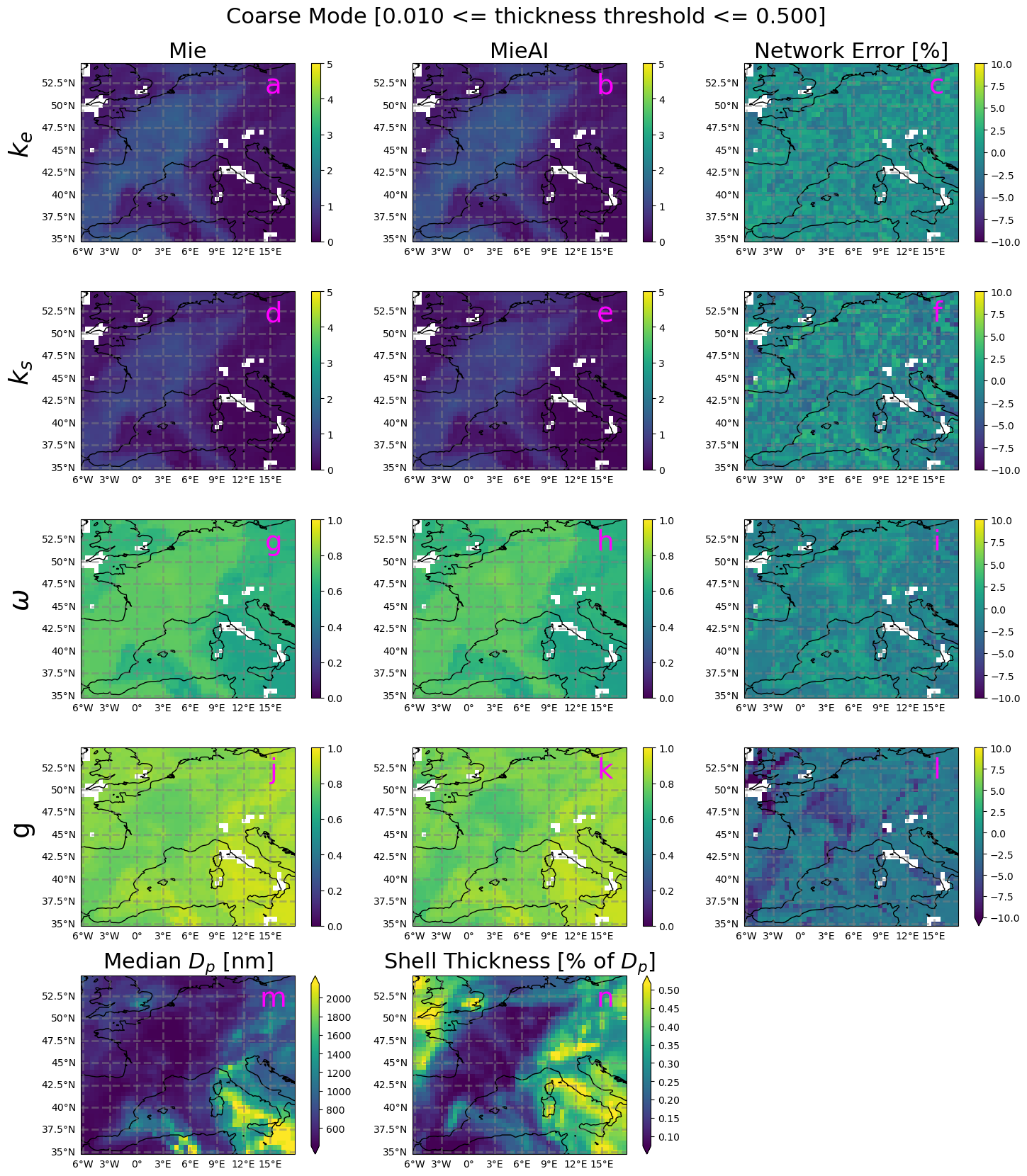}
    \caption{Same as Fig.~\ref{fgr:example7} but for coarse mode internally mixed aerosols at an altitude of 5 km above sea level in ICON-ART simulation of a dust event over central Europe.}
    \label{fgr:example9}
\end{figure*}

Next, we compare the AOP predictions of MieAI against the same estimated using Mie theory for the outputs of ICON-ART simulations and are shown in Figs. \ref{fgr:example7}, \ref{fgr:example8} and \ref{fgr:example9} for different real events examined in this study. For this purpose, the number concentrations for the constituent species of mixed modes were taken from ICON-ART output. We first map the ICON-ART number concentrations to RIs for core and shell as shown in Fig.~\ref{fgr:example41} and ICON-ART modes to bins assuming log-normal distributions as illustrated in Fig.~\ref{fgr:example2}. MieAI predictions for the bins are integrated back to modes and then compared with the Mie calculations.

Fig.~\ref{fgr:example7}  shows the spatial distribution of AOPs for simulated internally mixed volcanic aerosols in accumulation mode, obtained from both Mie and MieAI (see Fig.~\ref{fgr:s1} for the comparison in coarse mode). The illustration focuses on the derived AOPs after the La Soufrière volcanic eruption in April 2021, specifically showcasing the comparison at an altitude of 15 km above sea level 27 hours after the start of the simulation. The median diameters, shown in Fig.~\ref{fgr:example7}(m), exhibit a range spanning from 100 nm to 1200 nm. Concurrently, the shell (coating) thickness, depicted in Fig.~\ref{fgr:example7}(n), varies from 10 to 80\% of the total diameter. It is notable that a majority of particles possess median diameters exceeding 500 nm and exhibit thick coatings (more than 50\% coating fraction). In this case, volcanic ash constitute the core whereas water and inorganic species (sulphate, nitrate and ammonium) are the constituents of the coating/shell. A discernible trend emerges in the figure, where $Q_{sca}$ (Fig.~\ref{fgr:example7}(d)) and consequently $Q_{ext}$ (Fig.~\ref{fgr:example7}(b)) appear to align with the distribution of both median diameter and coating fraction. Higher values of $Q_{ext}$ and $Q_{sca}$ are observed in regions characterized by lower median diameters and coating fractions. Conversely, both $\omega$ (negative correlation; Fig.~\ref{fgr:example7}(g)) and $g$ (positive correlation; Fig.~\ref{fgr:example7}(j)) show a more pronounced correlation with changes in median diameter, with a lesser influence from the coating fraction. Impressively, MieAI, shown in Fig.~\ref{fgr:example7}(b, e, h, k), effectively captures these dependencies, showcasing an impressive agreement between its predictions and Mie theory estimates. The comparison between MieAI predictions and Mie theory estimates reveals a very good agreement, with relative errors (depicted in Fig.~\ref{fgr:example7}(c, f, i)) generally staying within 10\% for all AOPs, except for $g$ (Fig.~\ref{fgr:example7}(l)), where the relative error reaches up to 12\%. This suggests that the NN model effectively captures the intricate relationships between particle morphology, mixing state, and optical properties. Interestingly, it's worth noting that network errors exhibit a degree of dependency on the coating fraction for all AOPs, except $g$. In the case of $g$, network errors closely track the distribution of median diameter, with higher relative errors occurring in regions where the median diameter is smaller.

Fig.~\ref{fgr:example8} shows a comparison of the bulk AOPs estimated from MieAI and Mie  for case wildfire. This case study centers on an Australian wildfire event from 2019, specifically examining the comparison at an average altitude of 850 m above sea level after 23 hours of simulation (23rd of November 2019, 23:00 UTC). The selected altitude corresponds to the mass weighted height of the plume, further the plume at that level is wide spread with a high concentrations compared to other model levels. The time step selected is towards the end of the one day simulation, enabling transport and aging of the aerosol. In this case, soot constitutes the core whereas water, organic and inorganic species (sulphate, nitrate and ammonium) are the constituents of the coating/shell. Here, the median diameter (Fig.~\ref{fgr:example8}(m)) for the internally mixed aerosol in coarse mode ranges from 50nm to 1000nm whereas the shell (coating) thickness (Fig.~\ref{fgr:example8}(n)) varies from 35 to 50$\%$ of the total diameter (See Fig.~\ref{fgr:s2} for the comparison in accumulation mode). It's worth noting that this simulation predominantly features aerosol particles with total diameters exceeding 900 nm. Similar to case volcano, intriguing patterns emerge wherein all four optical properties exhibit alignment with the distribution of median diameters. In particular, changes in $Q_{ext}$ (Fig.~\ref{fgr:example8}(a)), $Q_{sca}$ (Fig.~\ref{fgr:example8}(d)), and $\omega$ (Fig.~\ref{fgr:example8}(g)) showcase a negative correlation with the variations in median diameters, while $g$ (Fig.~\ref{fgr:example8}(j)) demonstrates a positive correlation with the same. Intriguingly, none of the optical properties appear to exhibit sensitivity to variations in the shell thickness. Remarkably, the comparison between MieAI, shown in Fig.~\ref{fgr:example8}(b, e, h, k), and Mie calculations, shown in Fig.~\ref{fgr:example8}(a, d, g, j), underscores an excellent agreement, reaffirming the robustness of the employed NN model in effectively emulating Mie theory for internally mixed aerosols. The relative errors for all optical properties, shown in Fig.~\ref{fgr:example8}(c, f, i, l), in this case remain within the 10\%. Notably, in contrast to the La Soufrière case study, network errors in this instance appear to be particularly responsive to changes in median diameters rather than variations in the coating fraction.

Finally, Fig.~\ref{fgr:example9} shows a comparison of the MieAI predictions using the model trained with quantile transformation and Mie calculation for coarse mode internally mixed particles using ICON-ART simulation for case dust. The investigation focuses on a dust event occurring over central Europe, wherein the simulation encompasses a comprehensive range of aerosol species emissions, including sea salt, dust, and soot. The figure exclusively showcases these comparisons at an altitude of 5 km above sea level. In terms of particle characteristics, the median diameter (Fig.~\ref{fgr:example9}(m)) for mixed-phase aerosols within the coarse mode exhibits a range spanning from 200 nm to 2300 nm. Notably, the majority of these particles possess a median diameter of less than 500 nm. Concurrently, the shell (coating) thickness varies from 0 up to 50\% of the total diameter as shown in Fig.~\ref{fgr:example9}(n). However, it is important to note that a substantial proportion of the particles feature a shell thickness of less than 10\%. As anticipated, the optical properties (Fig.~\ref{fgr:example9}(a, d, g, j)) display sensitivity to changes in median diameter, mirroring the patterns observed in previous cases. While akin to the previous case, the influence of shell thickness remains relatively limited. As expected, MieAI (Fig.~\ref{fgr:example9}(b, e, h, k)) excels in capturing the variations in AOPs, with relative errors (Fig.~\ref{fgr:example9}(c, f, i)) staying below 10\% for $Q_{ext}$ and $Q_{sca}$ as well as $\omega$. The prediction accuracy for $g$ (Fig.~\ref{fgr:example9}(l)) is also reasonably strong, with errors generally remaining within 15\%. Importantly, it is noteworthy that the magnitude of errors in $g$ is sensitive to the coating fraction, a characteristic distinguishing it from the other three optical properties. For a complementary comparison in the accumulation mode, please refer to Fig.~\ref{fgr:s3}.

\begin{table*}[thbp]
\caption{Timing results (in seconds) of MieAI and Mie calculations for different real cases investigated in this study. Here, MieAI results are shown for the prediction batch size of 8192.}
\begin{tabular}{lcccc}
\tophline
Case & Number of grid cells & Mie & MieAI & Computational Gain \\
\middlehline

2021 La Soufrière Volcanic eruption & 73,500 & 423.3126 s & 0.2136 s & 1981.80x\\
2019--20 Australian Wildfire & 74,865 & 398.1323 s & 0.2108 s & 1888.67x\\
Summer 2019 dust event over central Europe & 28,800 & 132.9224 s & 0.1971 s & 0674.39x\\

\bottomhline
\end{tabular}
\label{tbl:2}
%\belowtable{} % Table Footnotes
\end{table*}

The comparisons between AOPs estimated using Mie theory and the predictions made by MieAI, employing a model trained without quantile transformation, are presented in Fig.~\ref{fgr:s4}. As clearly evident from the figure, the MieAI model, when trained without quantile transformation, exhibits notable shortcomings in capturing the variations in AOPs, with the exception of $Q_{ext}$. This discrepancy becomes particularly conspicuous despite the model's impressive performance on the test dataset, where correlation coefficients (R) exceeded 0.98 for all AOPs examined, including $Q_{ext}$, $\omega$, and $g$, as demonstrated in Fig.~\ref{fgr:s5}. 

The divergence between the model's performance on the test dataset and its application to real-world data underscores a critical limitation in its ability to generalize beyond the training context. Implications of this observation are far-reaching and offer valuable insights into the complexities of emulating intricate physical mechanisms using NNs, particularly when not validated against real-world scenarios. Consequently, it underscores the critical importance of comprehensive preprocessing of datasets before their integration into ML models, serving as a precautionary measure against potential pitfalls in model generalization.

In summation, this comprehensive analysis underscores the robustness of the MieAI model (with quantile transformation) in reproducing the optical properties of internally mixed aerosols. Note that the MieAI was trained using a dataset which had shell thickness up to 40\% only whereas the comparisons in all three cases include shell thickness beyond 40\% (up to 50\%). Thus, the comparisons clearly demonstrate the extrapolating capability of MieAI. The fact that the model successfully extrapolates its predictions beyond the training data's confines is a testament to its inherent capacity to generalize and capture the underlying physical principles governing the interactions between aerosol particles and light. This characteristic is particularly valuable in real-world scenarios where aerosol properties can exhibit a wide range of variability, often extending beyond the confines of training data. MieAI's capacity to accurately predict optical properties for aerosols with shell thicknesses up to 50\% highlights its versatility and reliability as a Mie emulator.

\subsection{Computational efficiency of MieAI}

In addition to its high fidelity in modeling AOPs, MieAI offers a remarkable advantage in computational efficiency, showcasing significant computational enhancements in comparison to traditional Mie calculations employed for the same purpose. As indicated in Table~\ref{tbl:2}, MieAI demonstrates a computational speedup exceeding 500 times that of Mie calculations across all scenarios investigated in this study.

The extent of performance gain is particularly noteworthy; for instance, during the 2019 dust event over central Europe with 28,800 ICON grid cells, MieAI exhibited a speedup of approximately 500 times. As the number of grids increases, this gain becomes more pronounced, with speedups surpassing three orders of magnitude when compared to Mie calculations. This phenomenon is exemplified in the ICON-ART simulations for events such as the La Soufrière volcanic eruption with 73,500 ICON grid cells, where MieAI achieved a speedup of around 1900 times, and the Australian wildfire event with 74,865 ICON grid cells, boasting a remarkable speedup of around 1800 times.

Furthermore, the computational cost associated with MieAI training is exceedingly minimal, taking approximately 3 hours and 20 minutes. This stands in stark contrast to the runtime requirements of ICON-ART simulations. Notably, MieAI training utilized a single computing node from a high-performance computing (HPC) cluster equipped with multiple nodes, each housing 36 Intel Xeon CPUs. It is pertinent to mention that both MieAI predictions and Mie calculations were executed utilizing a single CPU core.

\section{Discussion}

This study endeavors to introduce an innovative and computationally efficient framework, aptly named MieAI, specifically designed for calculating the bulk optical properties of internally mixed and coated aerosols characterized by a log-normal size distribution. Our approach leverages a straightforward multi-layer perceptron, a type of artificial neural network, to unravel the intricate relationship between AOPs and their physico-chemical characteristics, such as particle size distribution, mixing state, and chemical composition. Central to MieAI is the representation of both core and shell as ternary systems, subsequently linked to RIs via a volume mixing approach.

In order to validate the efficacy of our approach, we subjected it to rigorous evaluation against the gold standard method of Mie calculations -- a technique renowned for its precision albeit its notably sluggish computational speed. Our comparative evaluation unveiled that the NN-based MieAI approach not only attains remarkable accuracy -— with errors confined within 10\% -— but also exhibits an excellent computational efficiency, boasting a speed improvement of three orders of magnitude. 

Furthermore, our study underscores the paramount significance of meticulous pre-processing in enhancing the accuracy and generalizability of NN-based methodologies. We emphasize the necessity for rigorous evaluations of novel ML-based approaches prior to their widespread deployment in scientific applications. Moreover, MieAI model proposed in this paper tries to emulate the Mie calculations for thinly coated aerosols assuming the aerosols particles to be spherical and having the core-shell configuration. However, this approach can be extended to account for non-spherical shape and the morphologically complex configurations such as embedded, partly embedded, thick coating and partially embedded configurations \citep{Wang2023single, Riemer2019, Liu2018}.

With its generic design, the approach presented herein holds versatile applicability, seamlessly integrating into ACMs that adopt either bin or modal frameworks for representing aerosols and their optical properties. Moreover, the same framework can be extended to accommodate externally mixed aerosols and aerosol models featuring non-spherical shapes. 

The substantial precision achieved through our developed approach bears the potential to significantly contribute to the ongoing efforts aimed at mitigating uncertainties in aerosol forcing estimations. By bridging the gap between precision and computational efficiency, MieAI emerges as a valuable asset in the realm of physics-based weather and climate models, especially ACMs; poised to contribute substantially to advancing our understanding of aerosol-climate interactions and fostering more robust climate models.

%% The following commands are for the statements about the availability of data sets and/or software code corresponding to the manuscript.
%% It is strongly recommended to make use of these sections in case data sets and/or software code have been part of your research the article is based on.

\dataavailability{The training data and output from ICON-ART simulations generated in this study are available on \href{https://radar.kit.edu/radar/en/dataset/cMLguvOqjspbCtso}{Radar4KIT}.} %% use this section when having only data sets available

\codeavailability{The ICON model is openly available and is accessible through the following link: \href{https://icon-model.org/}{https://icon-model.org/}.  MieAI model and python codes used for performing analyses can be accessed here: \href{https://github.com/pankajkarman/MieAI}{https://github.com/pankajkarman/MieAI}.} %% use this section when having only software code available

%%\codedataavailability{} %% use this section when having data sets and software code available

%\sampleavailability{TEXT} %% use this section when having geoscientific samples available

%\videosupplement{TEXT} %% use this section when having video supplements available

%% REFERENCES

%% The reference list is compiled as follows:

\bibliographystyle{copernicus}
\bibliography{MieAI.bib}

\begin{acknowledgements}

 This work contributes to and is partly funded by the project \textit{PermaStrom} (grant no. 03EI4010A) within the seventh Energieforschungsprogramm of the German Federal Ministry of Economic Affairs and Climate Action (Bundesministerium für Wirtschaft und Klimaschutz, BMWK). This work used resources of the Deutsches Klimarechenzentrum (DKRZ) granted by its Scientific Steering Committee (WLA) under project ID 1070. Part of this research has been funded by the Deutsche Forschungsgemeinschaft (DFG) as part of the Research Unit VolImpact (FOR2820, DFG Grant 398006378). Open Access funding enabled and organized by Projekt DEAL.

\end{acknowledgements}

\authorcontribution{GAH conceived the idea. PK and GAH designed the research. PK wrote the NN code and performed the data analyses. GAH, JB, LMH, HK, and AR developed the ICON-ART code and carried out simulations. PK and GAH prepared the paper with significant contributions and comments on the original draft from all authors. GAH supervised the research at KIT.} %% this section is mandatory

\competinginterests{The authors declare no conflict of interest.} %% this section is mandatory even if you declare that no competing interests are present

%% \disclaimer{Publisher’s note: Copernicus Publications remains neutral with regard to jurisdictional claims in published maps and institutional affiliations. } %% optional section

%\clearpage

\appendix

\clearpage

%\subsection{} 

\appendixtables

\begin{table*}[t]
\caption{Various combinations (in percentage) of chemical species used to constitute the ternary systems for both core and shell to generate the dataset used for training MieAI.}
\begin{tabular}{lccc}
\tophline
& Constituent1 & Constituent2 & Constituent3 \\
\middlehline

1 & 100 & 0 & 0 \\
2 & 90 & 10 & 0 \\
3 & 90 & 0 & 10 \\
4 & 90 & 5 & 5 \\
5 & 80 & 20 & 0 \\
6 & 80 & 0 & 20 \\
7 & 80 & 10 & 10 \\
8 & 70 & 30 & 0 \\
9 & 70 & 0 & 30 \\
10 & 70 & 15 & 15 \\
11 & 60 & 40 & 0 \\
12 & 60 & 0 & 40 \\
13 & 60 & 20 & 20 \\
14 & 50 & 50 & 0 \\
15 & 50 & 0 & 50 \\
16 & 50 & 25 & 25 \\
17 & 40 & 60 & 0 \\
18 & 40 & 0 & 60 \\
19 & 40 & 30 & 30 \\
20 & 30 & 70 & 0 \\
21 & 30 & 0 & 70 \\
22 & 30 & 35 & 35 \\
23 & 20 & 80 & 0 \\
24 & 20 & 0 & 80 \\
25 & 20 & 40 & 40 \\
26 & 10 & 90 & 0 \\
27 & 10 & 0 & 90 \\
28 & 10 & 45 & 45 \\
29 & 0 & 0 & 100 \\
30 & 0 & 100 & 0 \\

\bottomhline
\end{tabular}
\label{tbl:s1}
\end{table*}

\begin{table*}[t]
\caption{Stage 1 of hyper-parameter tuning. Here, top 5 combinations of optimizer, number of neurons in a hidden layer, activation functions and number of hidden layers in the order of decreasing MSE values are shown.}
\begin{tabular}{lccccc}
\tophline
& optimizer & units & activation & number of layers & mse \\
\middlehline
1 &      adam &    64 &       gelu &          5 &  0.03209 \\
2 &      adam &    32 &       tanh &          3 &  0.03307 \\
3 &      adam &    64 &       tanh &          3 &  0.03372 \\
4 &      adam &    32 &       gelu &          4 &  0.03562 \\
5 &      adam &    64 &       gelu &          4 &  0.03632 \\
\bottomhline
\end{tabular}
\label{tbl:s2}
%\belowtable{} % Table Footnotes
\end{table*}

\begin{table*}[t]
\caption{Stage 2 of hyper-parameter tuning. Here, top 5 combinations of number of neurons in a hidden layer, learning rate, number of hidden layers and Minibatch size in the order of decreasing MSE values are shown.}
\begin{tabular}{lccccc}
\tophline
& units & learning rate & number of layers & batch size & mse \\
\middlehline
1 &   64 &  0.010 &         4 &       128 &  0.014102 \\
2 &   64 &  0.010 &         5 &       512 &  0.017238 \\
3 &   64 &  0.010 &         5 &       256 &  0.017359 \\
4 &   32 &  0.010 &         4 &       128 &  0.017747 \\
5 &   64 &  0.001 &         5 &       128 &  0.017767 \\
\bottomhline
\end{tabular}
\label{tbl:s3}
%\belowtable{} % Table Footnotes
\end{table*}

\clearpage

\appendixfigures

\begin{figure*}
\begin{subfigure}[t]{\linewidth}
    \centering\includegraphics[width=\linewidth]{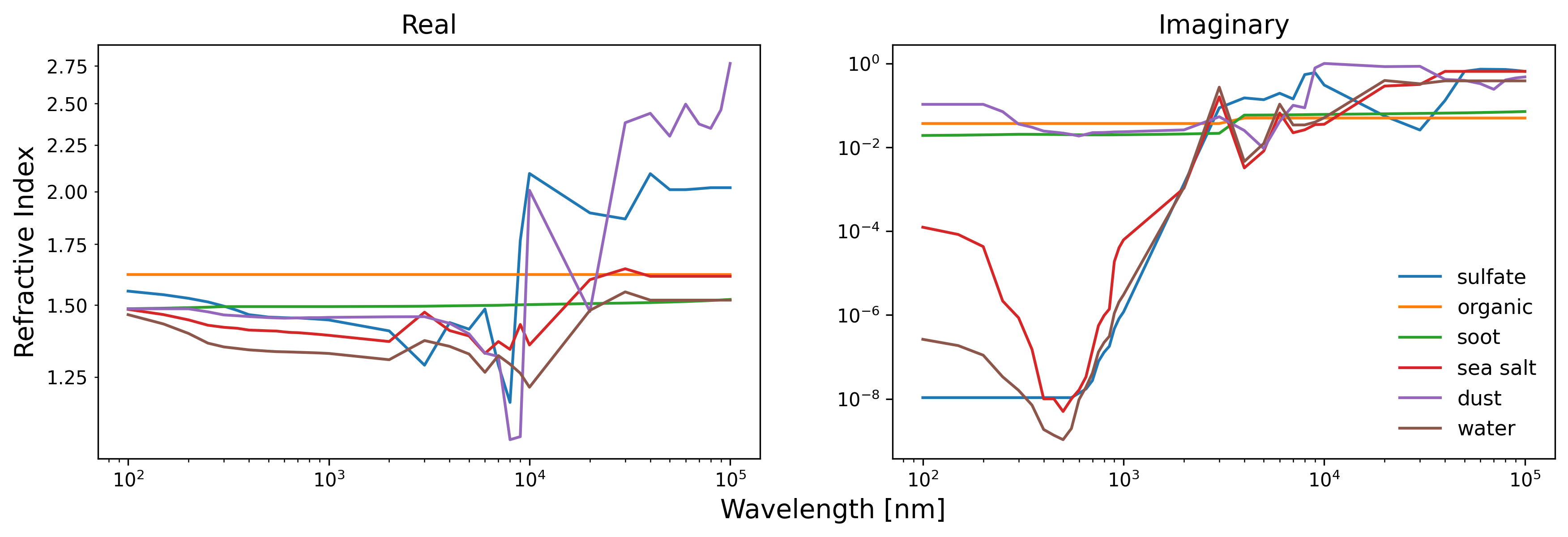}
    \caption{Refractive indices of the constituents of coated, internally mixed aerosol particles i.e. Mineral dust, sea salt, soot, water, inorganic and organic species.}
    \label{fgr:ri}
    \vspace*{4mm}
\end{subfigure}

\begin{subfigure}[b]{\linewidth}
   \centering\includegraphics[width=\linewidth]{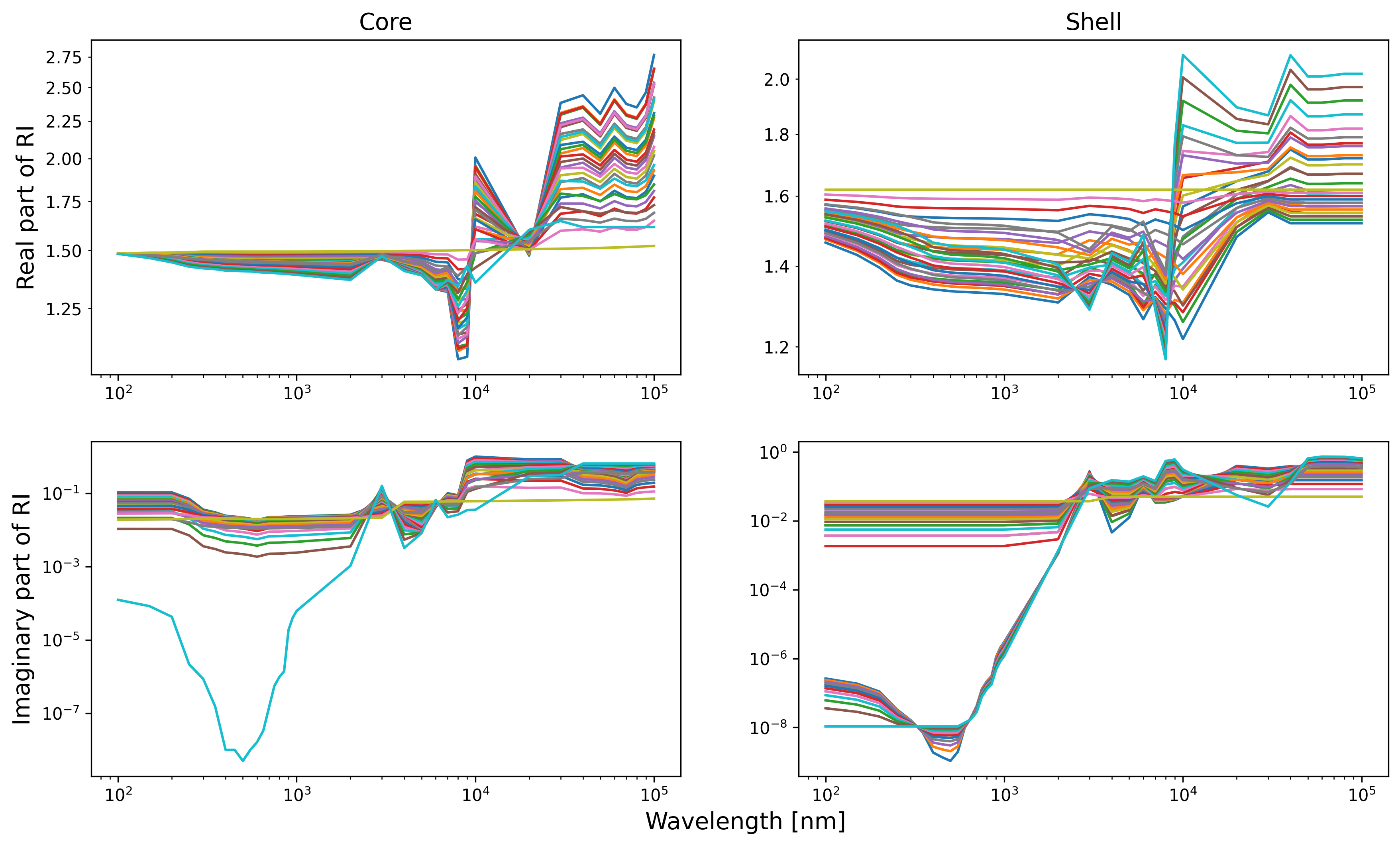}
    \caption{Real (upper row) and imaginary (lower row) part of refractive indices (RIs) for 30 different chemical compositions of core and shell. Here, the left column shows the RI for core and right column shows RI for shell.}
    \label{fgr:example41}
\end{subfigure}

\caption{Refractive indices of the internally mixed aerosol particle. Mineral dust, sea salt and soot constitute the core whereas water, Organic and Inorganic species constitute the shell of the particle.}
\label{fig:figures1}
\end{figure*}

\begin{figure*}[t]
    \includegraphics[width=\linewidth]{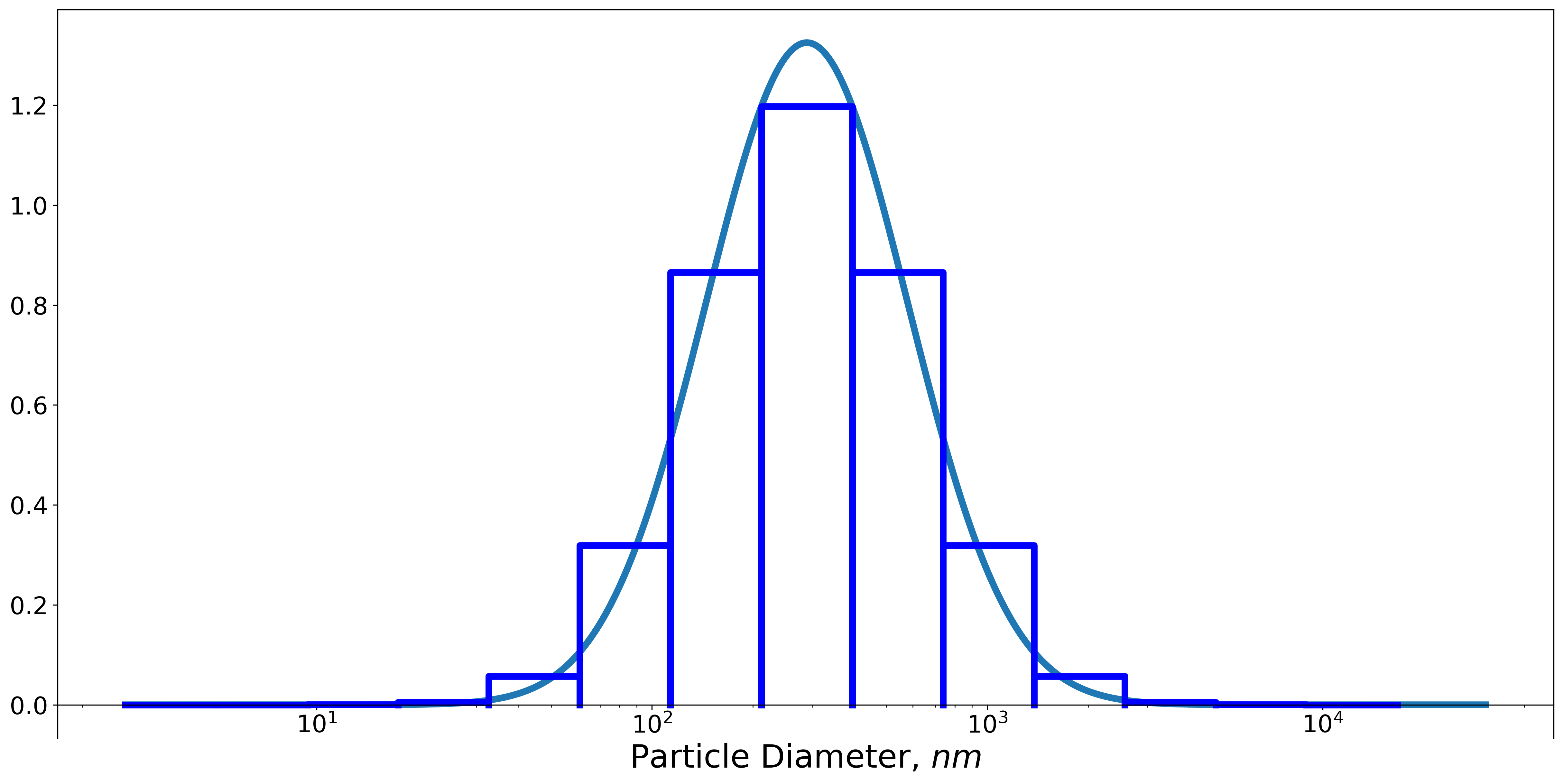}
    \caption{Mapping of aerosol mode to bins. Here, 15 log-normal bins are used to represent each aerosol mode.}
    \label{fgr:example3}
\end{figure*}

%\begin{figure*}[htbp]
%    \includegraphics[width=0.6\linewidth]{./figures/AOP.png}
%    \caption{Influence of shell thickness on optical properties of internally mixed aerosol particles. Upper panel shows the variation of mass extinction coefficient with wavelength whereas the lower panel shows the same for $\omega$. Here, blue line represents an aerosol particle with no soluble coating, the red one an internally mixed particle with 20\% shell thickness and the yellow one depicts particle with 40\% shell thickness. The total diameter of the aerosol particle is 5000 nm.}
%    \label{fgr:aop}
%\end{figure*}

\begin{figure*}[htbp]
    \centering\includegraphics[width=\linewidth]{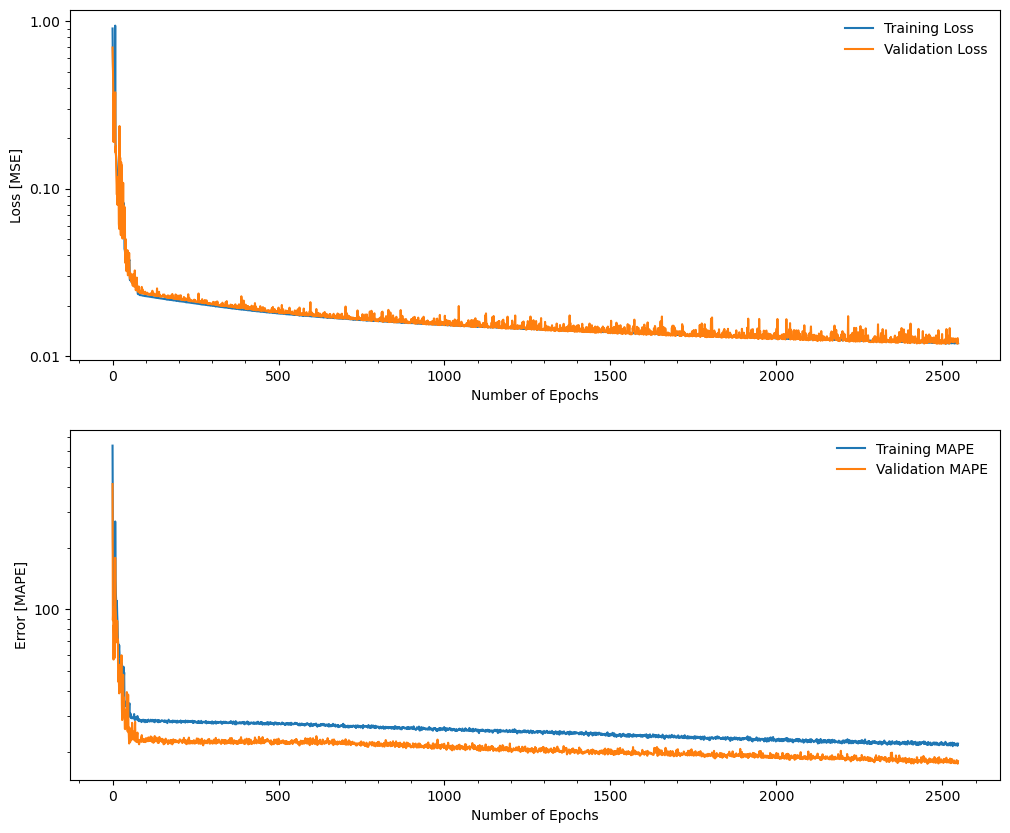}
    \caption{Variation of NN loss with epochs. Here, the green colored curve shows mean squared error (MSE) and mean absolute percentage error (MAPE) loss for the training samples whereas the orange colored curves represent the same for testing samples. Intended to train for 5000 epochs, the training of MieAI finishes at 2548 epoch due to earling stopping (patience parameter set to 50).}
    \label{fgr:example4}
\end{figure*}

\begin{figure*}[t]
    \includegraphics[width=\linewidth]{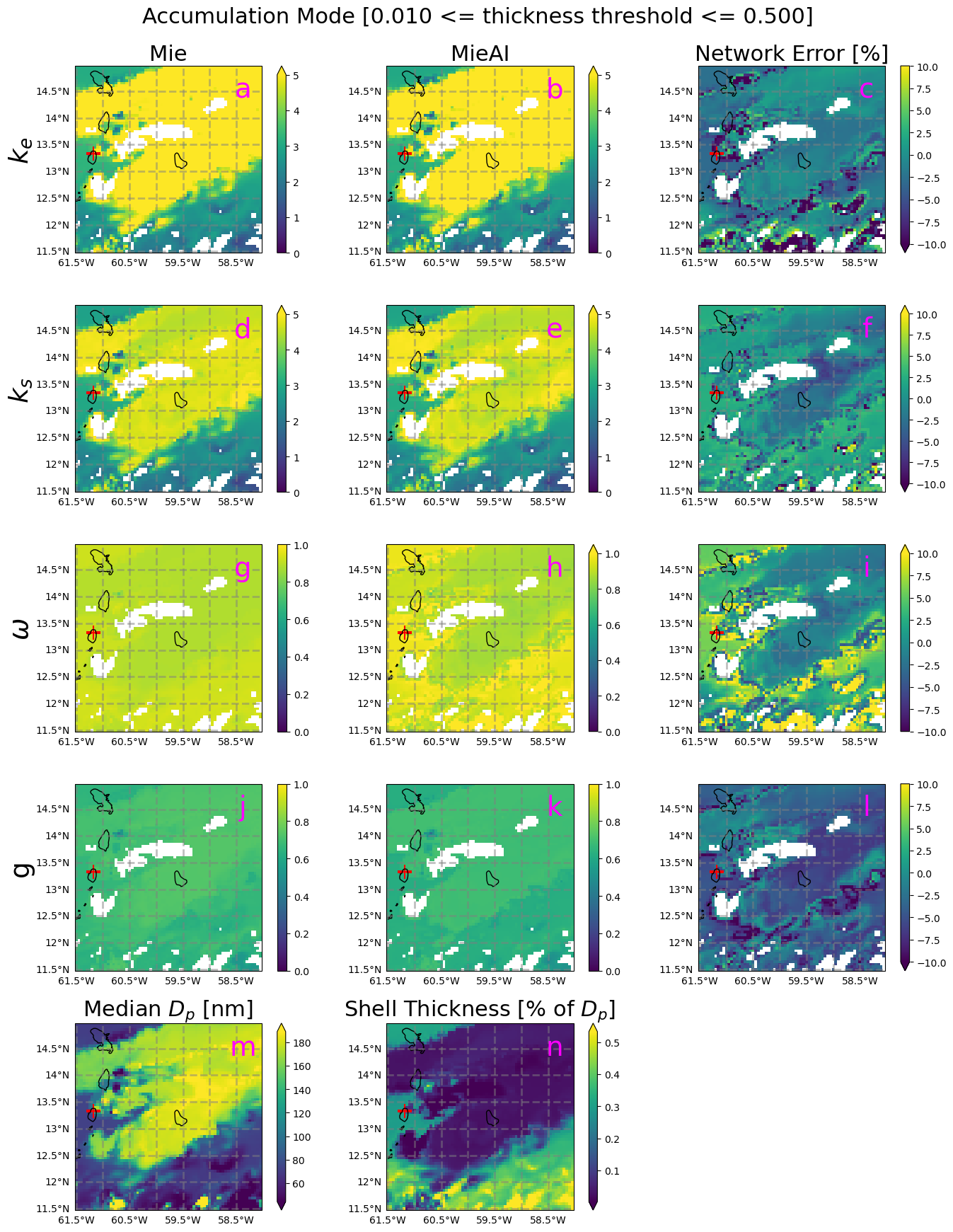}
    \caption{Same as Fig.~\ref{fgr:example7} but for accumulation mode internally mixed aerosols in ICON-ART simulation of the La Soufrière volcanic eruption event. Here, the plus symbol shows the location of volcanic eruption site.}
    \label{fgr:s1}
\end{figure*}

\begin{figure*}[t]
    \includegraphics[width=\linewidth]{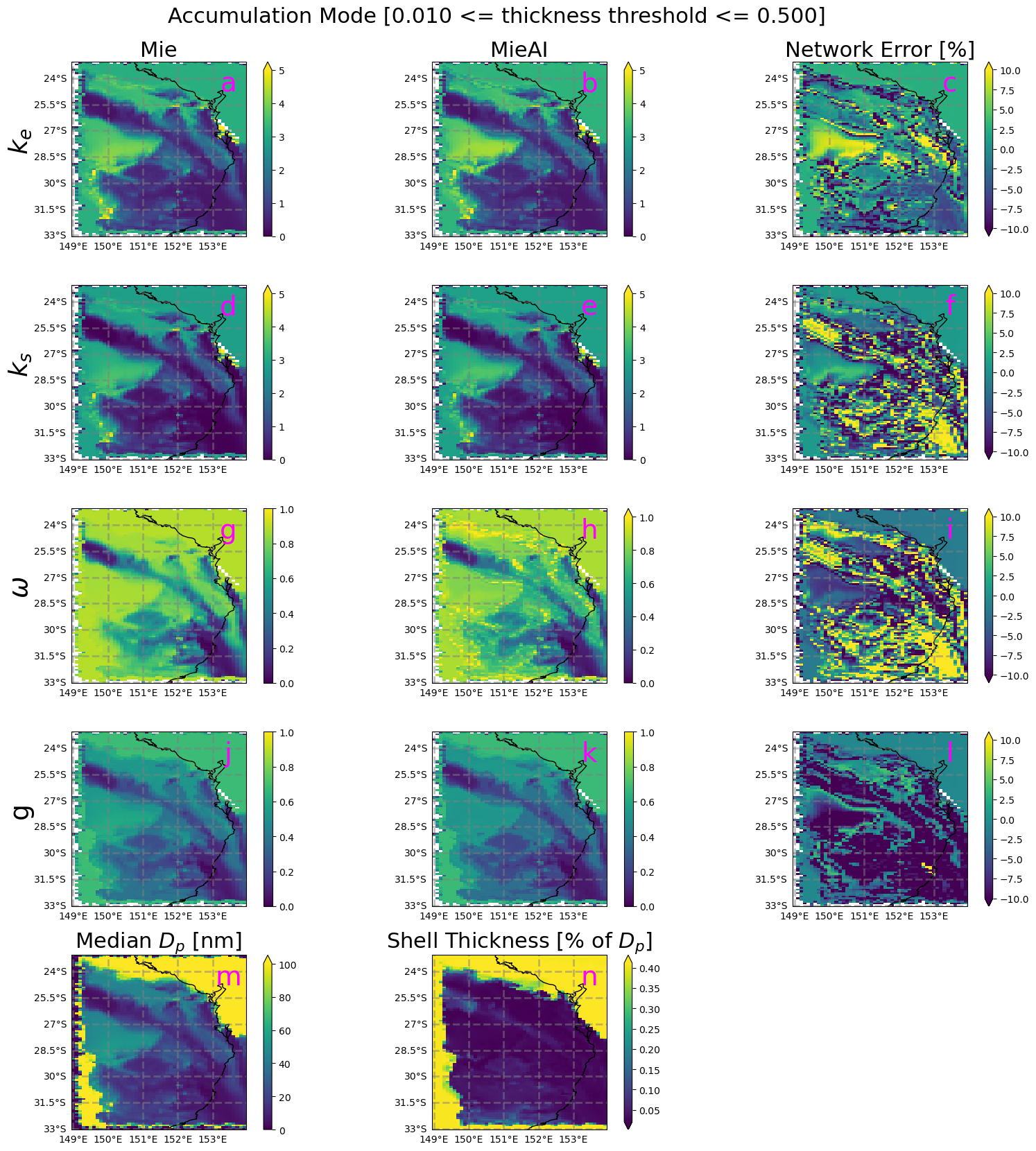}
    \caption{Same as Fig.~\ref{fgr:example7} but for accumulation mode internally mixed aerosols in ICON-ART simulation of Australian Biomass Burning event.}
    \label{fgr:s2}
\end{figure*}

\begin{figure*}[t]
    \includegraphics[width=\linewidth]{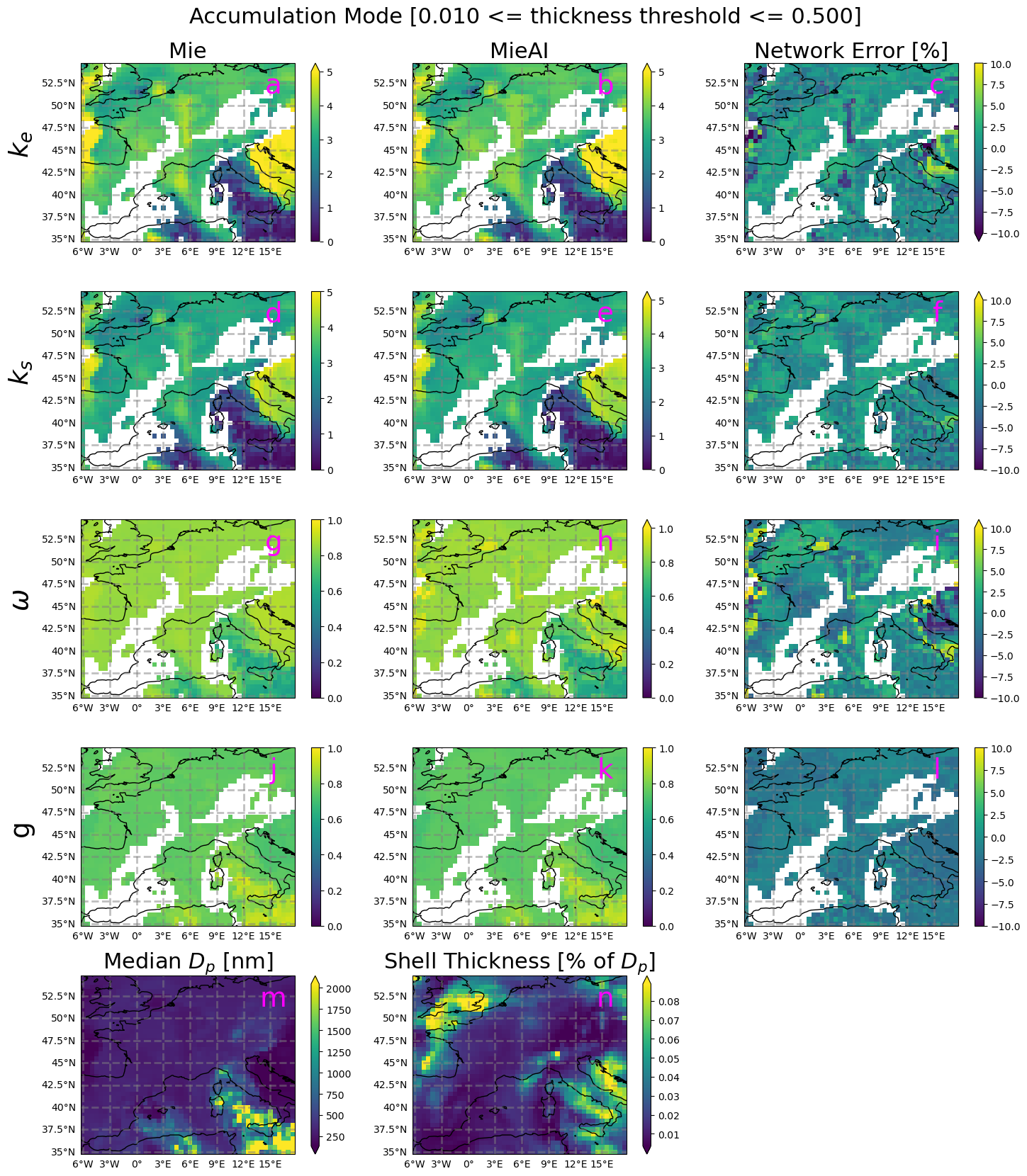}
    \caption{Same as Fig.~\ref{fgr:example7} but for accumulation mode internally mixed aerosols in ICON-ART simulation of a dust event over central Europe.}
    \label{fgr:s3}
\end{figure*}

%\section{MieAI predictions using the model trained without quantile mapping} 

\begin{figure*}[t]
    \includegraphics[width=\linewidth]{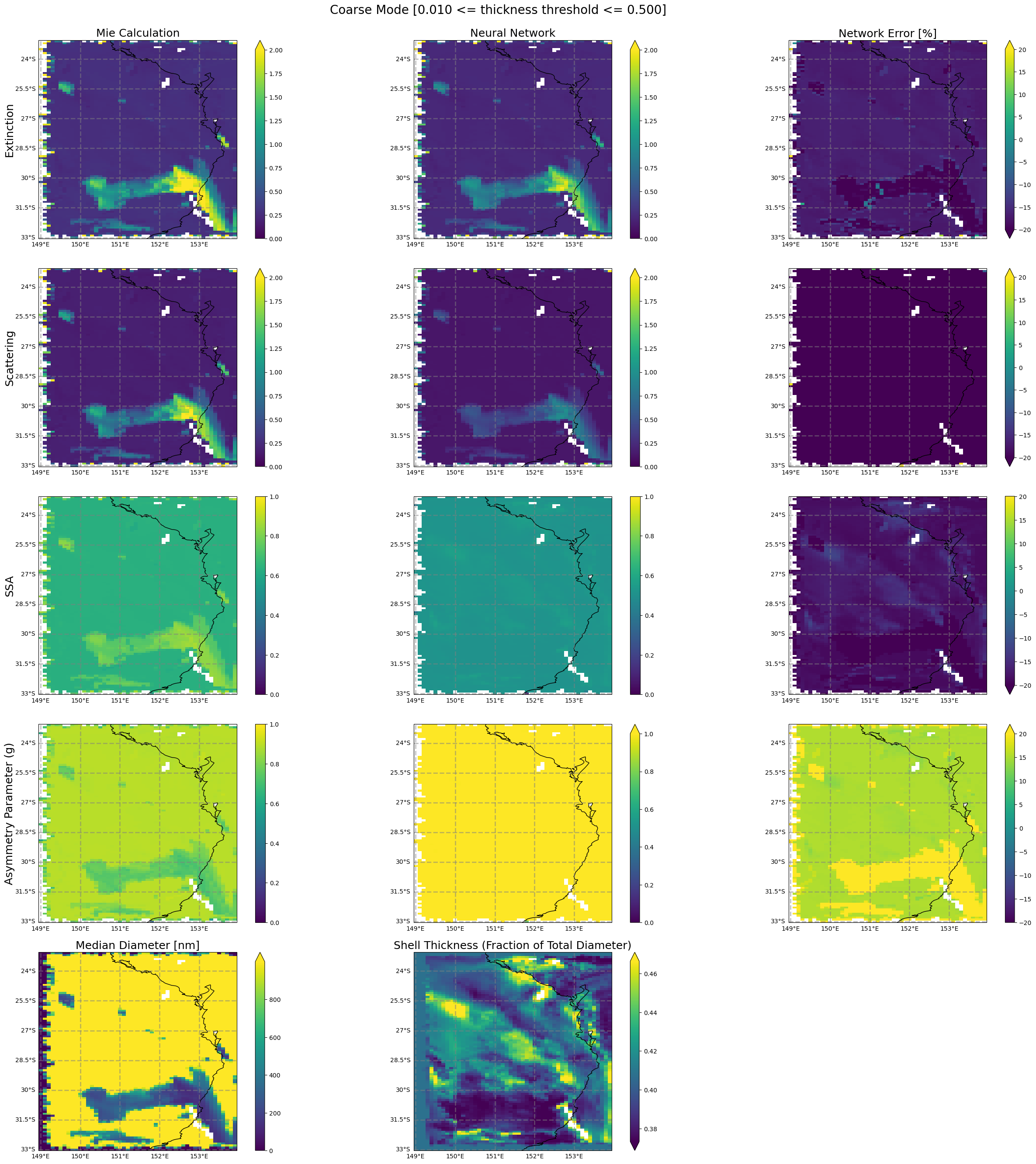}
    \caption{Same as Fig.~\ref{fgr:example7} but for ICON-ART simulation of Australian Biomass Burning event. Here, quantile mapping was not used for pre-processing the training data fed to MieAI.}
    \label{fgr:s4}
\end{figure*}

\begin{figure*}
\begin{subfigure}[t]{\linewidth}
    \centering\includegraphics[width=\linewidth]{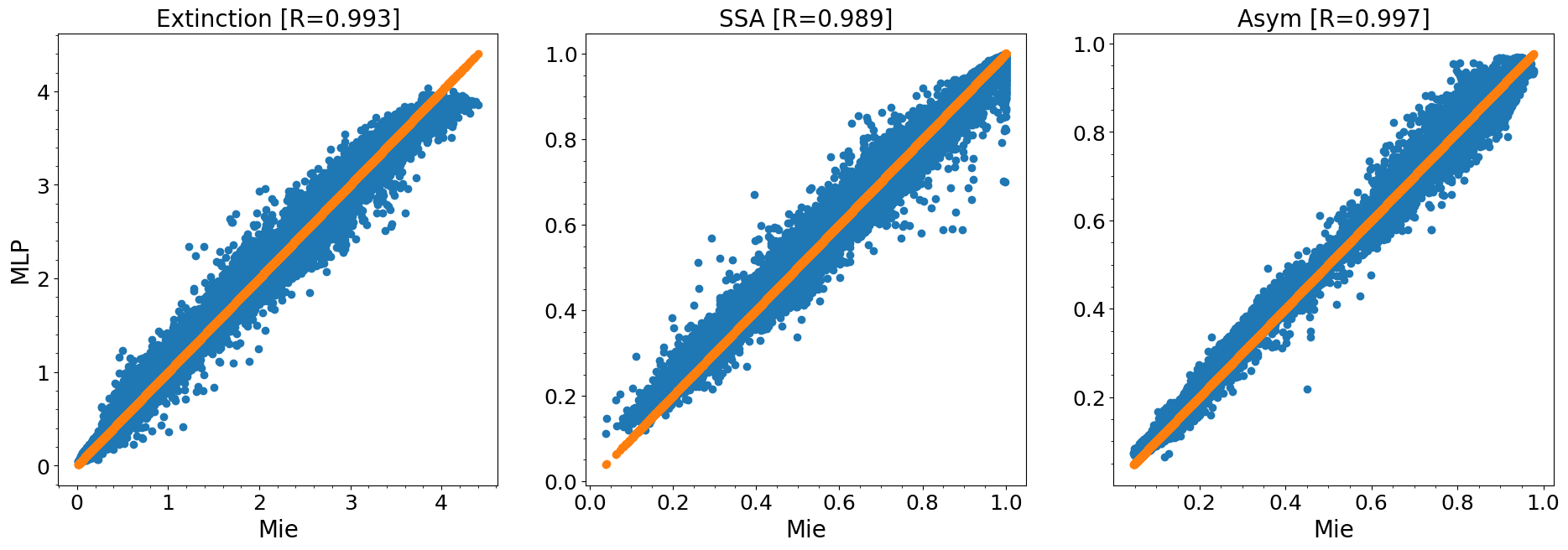}
    \caption{x >= 0.5}
    \label{fgr:x1}
    \vspace*{4mm}
\end{subfigure}

\begin{subfigure}[b]{\linewidth}
   \centering\includegraphics[width=\linewidth]{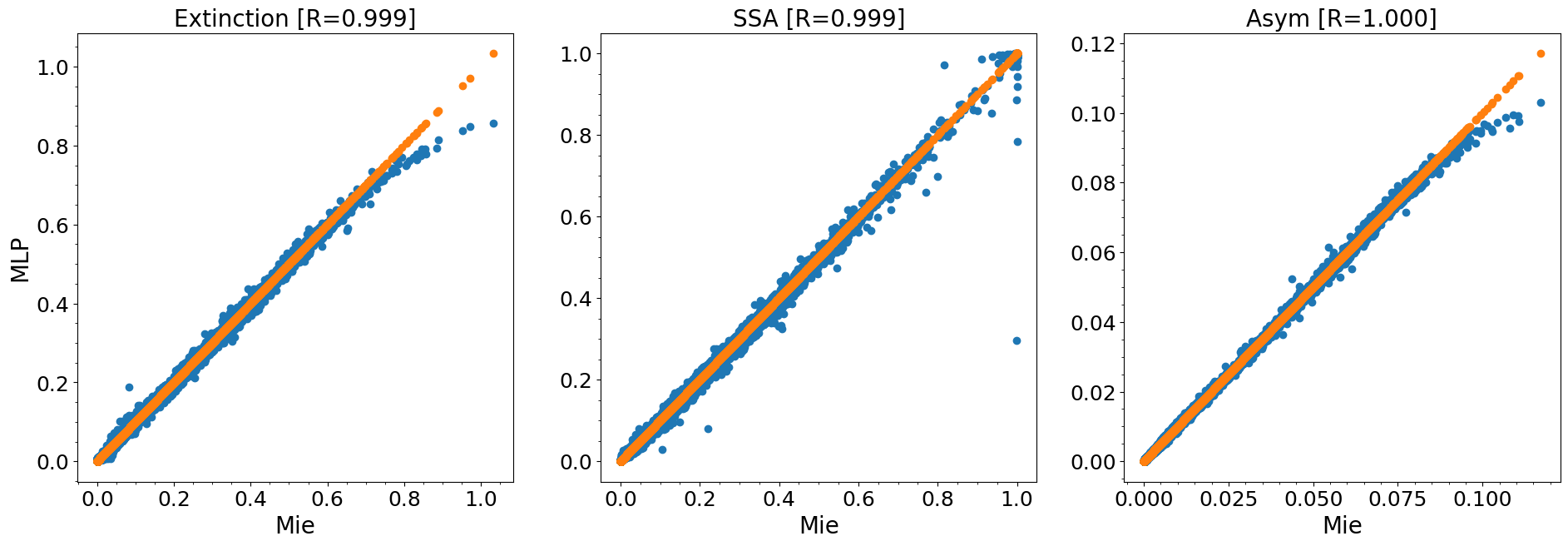}
    \caption{x < 0.5}
    \label{fgr:x2}
\end{subfigure}

\caption{MieAI predictions with model trained without quantile mapping against true AOPs estimated using Mie calculations for the test dataset. Here, we trained 2 separate NN models: one for size parameter upto 0.5 (Rayleigh scattering regime) and another for size parameter more than 0.5 (Mie Scattering regime).}
\label{fgr:s5}
\end{figure*}

%\section{}    %% Appendix A

%\subsection{}     %% Appendix A1, A2, etc.

\noappendix       %% use this to mark the end of the appendix section. Otherwise the figures might be numbered incorrectly (e.g. 10 instead of 1).

\end{document}